\documentclass[aps,prc,nofootinbib,twocolumn,showpacs]{revtex4-1}

\usepackage{multirow}
\usepackage{amsmath,epsfig,ulem}
\usepackage{amsfonts}
\usepackage{amsmath}
\usepackage{amssymb}
\usepackage{graphicx}
\usepackage{hhline,color}
\usepackage{pifont}
\usepackage{lmodern}
\usepackage[utf8]{inputenc}

\def\beq{\begin{equation}}
\def\eeq{\end{equation}}
\def\beqa{\begin{eqnarray}}
\def\eeqa{\end{eqnarray}}
\def\ban{\begin{eqnarray*}}
\def\ean{\end{eqnarray*}}
\def\bi{\begin{itemize}}
\def\ei{\end{itemize}}

\begin{document}

\title{Light Clusters and Pasta Phases in Warm and Dense Nuclear Matter}

\author{Sidney S. Avancini$^{1,2}$, M\'arcio Ferreira$^1$, Helena Pais$^1$, Constan\c ca
Provid\^encia$^1$, and Gerd R\"opke$^{3,4}$}

\affiliation{$^1$CFisUC, Department of Physics, University of Coimbra, 3004-516 Coimbra, Portugal. \\
$^2$Departamento de F\'{\i}sica, Universidade Federal de Santa Catarina, Florian\'opolis, SC, CP. 476, CEP 88.040-900, Brazil.\\
$^3$Institut f\"ur Physik, Universit\"at Rostock, D-18051 Rostock, Germany.\\
$^4$National Research Nuclear University (MEPhI), 115409 Moscow, Russia.}
\date{\today}

\begin{abstract}
The pasta phases are calculated for warm  stellar matter  in a framework of relativistic mean-field  models, including the possibility of light cluster formation. Results from three
  different semiclassical approaches are compared with a quantum statistical calculation. Light clusters are considered
as point-like particles, and their abundances are determined from the
minimization of the free energy. The couplings of the light-clusters
to mesons are determined from experimental chemical equilibrium
constants and many-body quantum statistical calculations. The effect of these light clusters on the chemical potentials is also discussed. It is shown that including heavy clusters, light clusters are present until larger nucleonic densities, although with smaller mass fractions.

\end{abstract}


\maketitle

%%%%%%%%%%%%%%%%%%%%%%%%%%%%%%%%%%%%%%%%%%%%%%%%%%%%%%%%%%%%%%%%%%%%%%%
%%%%%%%%%%%%%%%%%%%%%%%%%%%%%%%%%%%%%%%%%%%%%%%%%%%%%%%%%%%%%%%%%%%%%%%
\section{Introduction}

Presently, there is an increasing interest in the properties of warm and
dense matter in astrophysics and heavy ion physics,
i.e. nuclear/stellar matter at subsaturation densities (baryon density
$n_B \leq 0.15$ fm$^{-3}$) and moderate temperatures ($T \leq 20$
MeV). Light clusters seem to have an important role in the evolution
of core-collapse supernovae \cite{arcones08}, affecting in a non-negligible way the average energy of the electron anti-neutrinos. Also, in Refs. \cite{furusawa13,furusawa16}, it was shown that light clusters may influence in
a favorable or unfavorable way, depending on the conditions, the shock
revival in the  post-bounce phase of core-collapse supernovae.

The determination of light cluster abundances in warm and dense 
nuclear matter has been investigated using different approaches.
Recently, the formation of light clusters in low-density nuclear
matter, produced by heavy ion collisions (HIC), has been measured 
in the laboratory \cite{hagel12,qin12},
allowing the determination of quantities, such as in-medium
binding energies and chemical equilibrium constants. These and further
laboratory experiments give a strong evidence that in-medium
corrections are relevant for light clusters in nuclear matter at those densities and
temperatures. To obtain the corresponding nuclear equation of state
(EOS), a correct description of few-body correlations is essential.

The properties of warm dense matter are described by EOSs. 
In particular, for given constraints, such as the
temperature, $T$, and the number of neutrons and protons (densities
$n_n,\,\,n_p$, respectively), an ensemble in  thermodynamic
equilibrium, characterized by a thermodynamic potential, here the free
energy, can be defined. In stellar matter, allowing for weak
interaction processes,  $\beta$-equilibrium is established, and only
the baryon number (density $n_B=n_n+n_p$) can be chosen freely.
Other equations of state (thermodynamic, caloric, chemical potential, etc.) 
are derived from the thermodynamic potential in a consistent manner. 

Microscopically, the EOS can be derived within many-particle theory if
the interaction is known. However, approximations have to be
performed, and even the nucleon-nucleon interaction is not fully known,
in particular, in dense matter.
In the Brueckner approach, the nucleons in dense
matter are considered as quasiparticle states with momentum-dependent
energy shifts \cite{bhf}. In an alternative and simpler approach, 
the medium effects are introduced as semiempirical density functionals. 
Well-known examples are the  mean field approaches, like the Skyrme
parametrization  \cite{walecka,shf}
or the relativistic mean-field (RMF) models \cite{walecka,rmf,fsu,typel10}, which are fitted to reproduce 
the properties near the saturation density, see \cite{dutra1} and
\cite{dutra2} for recent compilations of Skyrme interactions and RMF
models, respectively.  As an important result, 
these mean-field approaches predict a phase transition in nuclear matter 
for sufficiently large proton fractions, $Y_p=n_p/n_B$.
Taking the Coulomb interaction into account, droplet formation (large nuclei), 
pasta structures, etc., are obtained, see, for instance, Ref. \cite{PCP15} 
and further references given there.

A drawback of a mean-field approach is that correlations are not directly
described, in particular, the formation of bound states. Correlations
become of importance at low temperatures and low densities. In the
nuclear statistical equilibrium (NSE) model, bound states (nuclei) are
considered as new components in addition to neutrons and protons, and
reactions bring the distributions of the respective components to
thermodynamic equilibrium as described by a mass action law. The
picture of an ideal mixture of components, which occasionally can
react if they collide, becomes, however, invalid for baryon densities
of the order of $10^{-3}$ fm$^{-3}$, or larger when mean-field modifications 
and the Pauli blocking are relevant. In particular, Pauli blocking suppresses 
the formation of light clusters, and at the Mott density, the clusters are dissolved \cite{RMS}.  

A quantum statistical (QS) approach can describe quantum correlations
in a systematic way. For instance, two-nucleon correlations and the
in-medium formation of deuteron and scattering phase shifts are given
in \cite{SRS}. The $\alpha$-like correlations are of particular
interest because of the relatively large binding energy of the
$\alpha$ particle. A quasiparticle concept can be worked out to
describe the light clusters ($d \equiv\, ^2$H,$\,\, t \equiv \,^3$H,
$\,\,h \equiv \,^3$He,$\,\,\alpha \equiv \, ^4$He) with binding
energies which depend not only on the center-of-mass momentum $\bf P$ relative to the medium,
but also on the parameters $T,n_B,Y_p$ characterizing the medium
\cite{roepke15}.

The light-nuclei quasiparticle approach has to take into account also 
the contribution of the continuum to reproduce the correct virial expansion 
for the thermodynamic quantities. Another problem exists when the spectral function, corresponding to the respective few-nucleon correlation function, 
shows no well developed peak structure so that the introduction of the 
quasiparticle approach becomes no longer well-defined.  This occurs, for instance, when the formation of larger clusters becomes relevant which leads to a background contribution to the spectral function,
but also at densities close to the saturation density, where the few-nucleon correlations are already implemented  in the mean-field contributions. The Mott effect reduces the contribution of light clusters so that a RMF approach is more adequate. More serious is the inclusion of larger clusters. Here, the QS description becomes too complex, and the replacement by semiempirical approaches, such as the Thomas-Fermi model, is necessary to get the correct physics. 

It is one goal of the present work to discuss the combination of 
light cluster approaches with pasta structure concepts. 
Light clusters (few-nucleon correlations) dominate at low densities and higher temperatures. 
In this density region of the phase diagram, the light clusters
determine the properties of nuclear matter. 
However, the inclusion of larger clusters using mean-field concepts is important 
if going to high densities, and a combination of both approaches is of interest. The combination of these light and heavy clusters has also been recently discussed in Ref. \cite{pais16}, where the authors used two different approaches, a generalized relativistic density functional and a statistical model with an excluded-volume mechanism, to compare the formation and dissolution of these aggregates in neutron star matter.  
Thus,  in this work, we focus on two questions: 
Are the chemical equilibrium constants, as derived from the abundances of light clusters, modified 
if the formation of droplets and pasta-like structures is taken into account?
How are the chemical potentials, calculated in a mean-field approach for stellar matter
with account of pasta-like structures, influenced if few-body correlations such as 
formation of light clusters are considered?

To combine QS calculations with RMF concepts, in recent works the light clusters are included in a
generalized RMF approach as additional degrees of freedom \cite{PCP15,typel10,ferreira12}. 
In particular, the effects of including light clusters in nuclear matter and the densities 
at which the transition between pasta configurations and uniform matter occur 
are investigated in \cite{PCP15}. As claimed there, more realistic parametrizations 
for the couplings of the light clusters should be implemented. 
The present work is aimed to contribute to this issue. 
For instance, the results obtained at low temperatures ($T=5, 10$ MeV) and low densities 
($n_B \approx 10^{-3}$ fm$^{-3}$) will be discussed. The goal is to find more precise data in this region.

The discussion about the necessity of including medium effects in the
EOS with the contribution of light clusters has emerged when the
chemical equilibrium constants (EC) were measured in heavy ion reactions \cite{qin12}. 
Definitively the nuclear statistical equilibrium (NSE) neglecting all in-medium effects was discarded. 
In addition to the QS approach to describe the chemical constants, 
the semiempirical excluded volume concept has been worked out further \cite{hempel2015}. 
Satisfactory agreement of excluded volume calculations with the QS method and the experimental data was found. 
In the present work, the approach which includes light clusters, as well as pasta phases, 
will be applied to the measured data for the chemical constants.

This paper is organized as follows. In Sec. \ref{sec:form}, we
  present the formalism for the calculation of matter including light clusters 
and pasta phases within a RMF approach. In Sec. \ref{sec:results}, 
some results are shown, and a comparison with experimental and QS results is made. 
Finally, in Sec. \ref{sec:conclusions}, a few conclusions are drawn.

%%%%%%%%%%%%%%%%%%%%%%%%%%%%%%%%%%%%%%%%%%%%%%%%%%%%%%%%%%%%%%%%%%%%%%%
%%%%%%%%%%%%%%%%%%%%%%%%%%%%%%%%%%%%%%%%%%%%%%%%%%%%%%%%%%%%%%%%%%%%%%%

\section{The Formalism} \label{sec:form}

In this section, we summarize the formalism that is used in this work. In particular, we review the RMF Lagrangian density, we discuss the way the cluster-meson couplings are fixed, and we present the density functional approach that has been applied to describe the pasta phases.

\subsection{Lagrangian}

We describe matter at subsaturation densities formed by protons, neutrons and
light clusters within a relativistic mean-field formalism \cite{typel10}. 
These particles interact through an isoscalar-scalar field $\phi$ with mass $m_s$,  an isoscalar-vector field $V^{\mu}$ with mass $m_v$, and an isovector-vector field  $\mathbf b^{\mu}$ with mass
$m_\rho$.  The light clusters  included in the calculation are the bosonic $\alpha$-particles and deuterons $d$, and the fermionic particles tritons ${}^3$H, represented by $t$, and  
helions ${}^3$He, represented by $h$.
A system of electrons with mass $m_e$ is also considered to  make matter neutral. 
The Lagrangian density of the system reads:
$$
\mathcal{L}=\sum_{j=n,p,t,h}\mathcal{L}_{j}+\mathcal{L}_{{\alpha }}+
\mathcal{L}_d+ \mathcal{\,L}_{{\sigma }}+ \mathcal{L}_{{\omega }} + 
\mathcal{L}_{{\rho }}
$$
\begin{equation}
 + \mathcal{L}_{\omega \rho} + \mathcal{L}_e +\mathcal{L}_A.
\label{lag}
\end{equation}
where the term $\mathcal{L}_{j}$ is given by 
\begin{equation}
\mathcal{L}_{j}=\bar{\psi}_{j}\left[ \gamma _{\mu }iD^{\mu }_j-M^{*}_j\right]
\psi _{j}  \label{lagnucl},
\end{equation}
and the $\alpha$ particles and the deuterons are described as in 
\cite{typel10}, with $\mathcal{L}_{{\alpha }}$ and $\mathcal{L}_d$ given, respectively, by
\begin{equation}
\mathcal{L}_{\alpha }=\frac{1}{2} (i D^{\mu}_{\alpha} \phi_{\alpha})^*
(i D_{\mu \alpha} \phi_{\alpha})-\frac{1}{2}\phi_{\alpha}^* {M^*_{\alpha}}^2
\phi_{\alpha},
\end{equation}
and
$$ \mathcal{L}_{d}=\frac{1}{4} (i D^{\mu}_{d} \phi^{\nu}_{d}-
i D^{\nu}_{d} \phi^{\mu}_{d})^*
(i D_{d\mu} \phi_{d\nu}-i D_{d\nu} \phi_{d\mu})$$
\begin{equation}
-\frac{1}{2}\phi^{\mu *}_{d}{ M^*_{d}}^2
\phi_{d\mu},
\end{equation}
with 
\begin{eqnarray}
iD^{\mu }_j &=&i\partial ^{\mu }-g_{vj}V^{\mu }-g_{\rho j }
{\boldsymbol{t}}%
\cdot \mathbf{b}^{\mu } - q_{ei} A^\mu
, \label{Dmu}
\end{eqnarray}
 $ j=n,p,t,h,\alpha,d$, where  $\boldsymbol {t}$ stands for the isospin operator,
 $M^*_i$ is the effective mass, $g_{vi}$, and $g_{\rho i}$ are
   the particle $i$-meson couplings, and $q_{ei}$ is the electric charge of particle $i$. They are defined in the next section.

The meson and  photon contributions in eq. (\ref{lag}) are given by
\begin{eqnarray}
\mathcal{L}_{{\sigma }} &=&\frac{1}{2}\left( \partial _{\mu }\phi \partial %
^{\mu }\phi -m_{s}^{2}\phi ^{2}-\frac{1}{3}\kappa \phi ^{3}-\frac{1}{12}%
\lambda \phi ^{4}\right)  \\
\mathcal{L}_{{\omega }} &=&\frac{1}{2} \left(-\frac{1}{2} \Omega _{\mu \nu }
\Omega ^{\mu \nu }+ m_{v}^{2}V_{\mu }V^{\mu }
+\frac{1}{12}\xi g_{v}^{4}(V_{\mu}V^{\mu })^{2} 
\right) \\
\mathcal{L}_{{\rho }} &=&\frac{1}{2} \left(-\frac{1}{2}
\mathbf{B}_{\mu \nu }\cdot \mathbf{B}^{\mu
\nu }+ m_{\rho }^{2}\mathbf{b}_{\mu }\cdot \mathbf{b}^{\mu } \right)\\
\mathcal{L}_{\omega \rho } &=& \Lambda_v g_v^2 g_\rho^2 V_{\mu }V^{\mu }
\mathbf{b}_{\mu }\cdot \mathbf{b}^{\mu }\\
\mathcal{L}_{A} &=&-\frac{1}{4} F_{\mu \nu }F ^{\mu \nu }~,
\end{eqnarray}
where $\Omega _{\mu \nu }=\partial _{\mu }V_{\nu }-\partial _{\nu }V_{\mu }$, 
$\mathbf{B}_{\mu \nu }=\partial _{\mu }\mathbf{b}_{\nu }-\partial _{\nu }
\mathbf{b}
_{\mu }-g_{\rho }(\mathbf{b}_{\mu }\times \mathbf{b}_{\nu })$ and $F_{\mu \nu }=\partial _{\mu }A_{\nu }-\partial _{\nu }A_{\mu }$.

Electrons will be included in stellar matter, with  the electron Lagrangian density  given by
\begin{equation}
\mathcal{L}_e=\bar \psi_e\left[\gamma_\mu\left(i\partial^{\mu}+ e A^{\mu} 
\right)-m_e\right]\psi_e.
\label{lage}
\end{equation}

The  parameters $\kappa$, $\lambda$ and $\xi$ are
self-interacting  couplings and the  $\omega-\rho$ coupling
$\Lambda_v$ is included to soften the density dependence of the
symmetry energy above saturation density. 
In the present study, we  always consider the FSU model \cite{fsu}. 
Values for the parameters $\kappa$, $\lambda$ and $\xi$, but also for the coupling constants
and the masses of the mesonic  components, are given in Refs. \cite{fsu,ferreira12}.
The contribution of the hadronic components are discussed in the following section.

\subsection{Medium modified masses of the hadronic components}

The treatment of warm and dense nuclear matter, including light cluster and pasta phases,
demands the appropriate treatment of nucleons in a dense medium. Different approaches are 
possible, and have been extensively investigated for the single nucleon ($n,p$) contribution.
Within a QS approach, a spectral function can be deduced. Then, the quasiparticle concept may 
be introduced, where the energies of the nucleons are shifted because
of medium effects.  Note, however, that  heavy clusters have never
  been included in this approach, and will not be considered in
the present study.
Results of microscopic calculations, such as Brueckner (DBHF) calculations, can be represented by
RMF models, which contain parameters adapted to known data, e.g. the properties of nuclei and 
nuclear matter near the saturation density.

For the single nucleon contribution, within the RMF approach, we have the density-dependent effective mass
\begin{eqnarray}
M^{*}_j &=&M^*=M -g_{s}\phi, \quad j=n,p .
\label{Mstar}
\end{eqnarray}
 We consider the same  mass for protons and neutrons in
the spirit of the RMF model proposed by Walecka
\cite{walecka}. We do not expect that for the present calculation
at finite temperature and large proton fractions this approximation
has a noticeable effect. However, for subsaturation cold stellar  matter in
$\beta$-equilibrium, finite effects may be expected and the experimental masses
should be adopted, which may be done in a straightforward manner. 
Together with (\ref{Dmu}), the quasiparticle shift of the nucleons is
described within the RMF approach.
 From a more general point of view, the  RMF approach used to describe  warm
and dense matter can be seen as an effective field theory built in the
framework of a density functional theory, where the many-body effects
are included in the parameters of the model.

The inclusion of correlations, in particular the formation of light clusters, is a delicate problem in the RMF approach.
The calculation of the few-body spectral function from which in-medium correlations, in particular bound state formation,
are derived, is subject of a QS approach. In full analogy to the concept of single-nucleon quasiparticles, bound states 
which appear as poles of the few-body spectral functions, can be considered as quasiparticles, with medium-modified 
energies.

Within the QS approach,  this medium modification of the binding energy of nuclei has two reasons. Firstly, the self-energy shift of the constituting nucleons gives a shift of the quasiparticle energy of clusters which is treated in the same manner as the 
quasiparticle shift of single-nucleon quasiparticles. Secondly, the Pauli blocking due to the surrounding medium produces 
a shift of the binding energy which, in contrast to the single-nucleon quasiparticle shift, is strongly dependent on
temperature and center-of-mass momentum of the bound state. The strong decrease of the binding energy of nuclei,
because of Pauli blocking, leads to the dissolution of light clusters already at low nucleon densities. However, the disappearance of bound states with increasing density is not an abrupt change of the properties because the bound states with large c.o.m. momentum can survive up to higher densities,  so that the correlations representative for light clusters are present also at higher densities and only smoothly disappear.
 
Similar to the single-nucleon quasiparticles $\{n,p\}$, the
light-cluster quasiparticles $\{d,t,h,\alpha\}$ are considered as
additional degrees of freedom in the Lagrangian (\ref{lag}). The
coupling of clusters to the meson fields should reproduce the shift of the corresponding quasiparticle energies.
We have in the low-density limit, where Pauli blocking effects can be neglected,
\begin{eqnarray}
M^{*}_i &=&M_{0i} -g_{si}\phi,  \\
M_{0i} &=&A_iM - B_{0i}, \quad i=d,t,h,\alpha,
\label{Mstar1}
\end{eqnarray}
where $B_{0i}$ are the binding energies of the particles in the vacuum,
$B_{0d}=2.224$ MeV, $B_{0t}=8.482$ MeV, $B_{0h}=7.718$ MeV, and
$B_{0\alpha}=28.296$ MeV.  For the average vacuum nucleon mass, we take the value $M = 939$ MeV. 
 
To include the Pauli blocking shift
\cite{roepke11,typel10}, dependent on the c.o.m. momentum, temperature and density, 
we improve previous approaches \cite{ferreira12,PCP15}. 
As in the case of the nucleons $\{n,p\}$, where the coupling constants are fitted to describe known 
properties of nuclei and nuclear matter, we need experimental data or
first principles theoretical calculations  to determine the cluster-meson coupling parameters.
Results for the  properties of nuclear matter at low densities are
still missing. A benchmark is obtained from the virial expansions
\cite{SRS,HS} and an interesting result that gives some information
about  the medium modifications of light clusters at low densities 
are the chemical equilibrium constants (EC)  which  have been
calculated in \cite{qin12}.

In the present approach, we are going to model  medium effects with an appropriate choice of the cluster-meson
couplings, where the binding energy of the cluster in the
medium is defined by
\begin{equation}
B_i=A_i M^*-M^*_i.
\label{Bi}
\end{equation}
 The couplings are written as $g_{sj}=x_{sj} g_s$,   $g_{vj}= x_{vj} g_v$
and $g_{\rho j}=|Z_j-N_j| g_{\rho}$, 
where $A_j$ is the mass number, 
$Z_j$  the proton number, and 
$N_j$ the neutron number. 
The parameters $x_{ij}$ are fixed in the following way:
a) $x_{sj}=x_{vj}=1$, for $j=p,n$; b) 
 $x_{si}=\frac{3}{4}A_i$, for
 $i=d,t,h,\alpha$, as proposed in Ref. \cite{ferreira12}, because these
 parameters reproduce quite well the binding energy given in Ref.
 \cite{typel10}, for $T=5$ MeV, and the experimental predictions of the  Mott
 densities at $T=5$ MeV, given in Ref. \cite{hagel12};
c)  the parameters $x_{vi}$  are fixed as in \cite{ferreira12}, so that the
dissolution density at $T=0$ of each type of cluster, defined as the density at which the free energy of
clusterized matter equals the free energy of nucleonic matter,  is the one obtained in \cite{typel10}, where
a statistical approach was used. For the FSU \cite{fsu} EoS, these
  $x_{vi}$ ratios are given by, see  \cite{ferreira12},
\begin{equation}
\left(\begin{array}{c} x_{vd}\\x_{vt}\\x_{vh}\\x_{v\alpha}
\end{array}\right)= 
\left(\begin{array}{c} 3.516\\4.382\\4.624\\5.675
\end{array}\right)
\eta
\label{gveta}
\end{equation}
with $\eta=1$.  In the present work, we will allow $\eta$ to vary,
in order to be able to reproduce the experimental EC. We point out
that, in principle, we should consider different values of $\eta$ for
the different clusters, but we have avoided this approach to keep the
parameter space restricted. We postpone an overall optimization of
all the parameters for a  future work;
d) for the coupling to the $\rho$-meson we consider the simplest
approach and take the  coupling  proportional to the
isospin projection of the light cluster. A more realistic choice could
be done once the couplings to the $\sigma$ and $\omega$ mesons are
more constrained.

Note that the coupling constant, $g_v$, describes the repulsive interaction because of Pauli blocking. This is the case for the nucleon-nucleon interaction, where the Pauli blocking acts on the quark substructure of nucleons; it is only weakly dependent on $T$.
For the medium shift of the binding energy of light nuclei, also the Pauli blocking is responsible, but because of the different energy scale
of the binding energies, the dependence on $T$ is strong.
 An effective field theory, that takes  into account this temperature dependent Pauli
blocking effect,  will  require temperature dependent
parameters as implemented in \cite{typel10}, and  leads to
 more complex thermodynamics. In the present
study, we have kept to constant couplings, and, therefore, 
the fit we are performing (e.g. $\eta =0.7$)
 is valid for the temperature region under consideration (5-10 MeV)
 but cannot be taken from $T = 0$ MeV (where $\eta = 1$).

\subsection{Density functional approach}

The calculation of the warm pasta phase, including light clusters, i.e.,
$\{ n, p, d, t, h,\alpha, e\}$ matter, is 
performed  using the numerical prescription given in
\cite{pasta1}. In this approach, the fields are assumed to vary slowly so that
the nucleons and the clusters can be treated as moving in locally constant fields at each point.
The finite temperature semiclassical Thomas Fermi (TF)  approximation
is obtained within a density functional formalism.
We start from the grand canonical potential density:
\begin{eqnarray}
 {\omega } &=& \omega ( \{f_{i+}\},\{f_{i-}\}, \{F_j\}, \{\nabla F_j\}
               )\nonumber\\
& =&
{\cal E}_t-T{\cal S}_t-\sum_{i=p,n,e,d,t,h,\alpha}\mu_i \rho_i ~ ,~
\label{grand}
\end{eqnarray}
where  $\{f_{i+}\}$, $\{f_{i-}\}$, $i=p,n,e,d,t,h,\alpha$  stand for the  proton,
neutron, electron, light clusters and respective
anti-particle distribution functions, defined in Eq. (\ref{dist}), 
and $ \{F_j\} $, $ \{\nabla F_j\}$  represent the fields $\phi_0,\,  V_0,
\, b_0, \,A_0$ and respective gradients.
The quantities
 ${\cal E}_t = {\cal E} +{\cal E}_e $  and ${\cal S}_t = {\cal S}
 +{\cal S}_e$ are
the total energy and entropy densities,
respectively. The total energy density  is a functional of the density, and was defined in \cite{pasta1} for
$T=0$. For finite temperatures, we have a similar expression
\begin{equation}
{\cal E}_{t}=
\sum_{i=p,n,e,d,t,h,\alpha} E_i(\mathbf r) +g_v V_0(\mathbf r)\rho_v(\mathbf r) + g_\rho b_0(\mathbf r) \rho_3 (\mathbf r)
 \nonumber 
\end{equation}
\begin{equation}
+\frac{1}{2} \left[(\nabla \phi_0(\mathbf r))^2 +  
m_s^2\phi_0^2(\mathbf r)\right]   
+\frac{\kappa}{3!} \phi_0^3(\mathbf r) 
+\frac{\lambda}{4!} \phi_0^4(\mathbf r) 
\nonumber 
\end{equation}
\begin{equation}
-\frac{1}{2}\left[(\nabla V_0(\mathbf r))^2 +m_v^2 V_0^2(\mathbf r)
+\frac{\xi g_v^4}{12} V_0^4(\mathbf r)\right]
\nonumber 
\end{equation}
\begin{equation}
 -\frac{1}{2} \left[(\nabla b_0(\mathbf r) )^2 
+m_{\rho}^2 b_0^2(\mathbf r)\right]-\Lambda_v\,g_v^2\,V_0^2(\mathbf r)\,g_\rho^2\, b_0^2(\mathbf r)
\end{equation}
\begin{equation}
-\frac{1}{2}\left[\nabla A_0(\mathbf r) \right]^2 + e\rho_q(\mathbf r)A_0(\mathbf r) ~~,
\label{enfun} 
\end{equation}
where 
\begin{eqnarray}
E_i=&&\frac{\gamma_i}{2\pi^2}\int dp\, p^2
  \sqrt{p^2+{M_i^{*}(\mathbf r)}^2}\left(f_{i+}(\mathbf r, \mathbf
  p)+f_{i-}(\mathbf r, \mathbf p)\right) ,\nonumber\\
&&i=p,n,d,h,t,\alpha~~, 
\end{eqnarray}
with $\gamma_i=2s_i+1$, the spin degeneracy of particle $i$, and 
\begin{equation}
E_e=\frac{1}{\pi^2}\int dp\, p^2 \sqrt{p^2+m_e^2}\left(f_{e+}(\mathbf r, \mathbf p)+f_{e-}(\mathbf r, \mathbf p)\right) ~~.
\end{equation}
In the above expressions,
\begin{equation}
\rho_v{(\mathbf r)}= \sum_{i=p,n,d,t,h,\alpha} {x_{vi}} \rho_i(\mathbf r),
\end{equation}
\begin{equation}
\rho_3{(\mathbf r)}= \sum_{i=p,n,t,h} t_{3i}\rho_i(\mathbf r),
\end{equation}
and
\begin{equation}
\rho_q{(\mathbf r)}= \sum_{i=p,d,t,h,\alpha} \frac{q_{ei}}{e}\rho_i(\mathbf r),
\end{equation}
with
\begin{equation}
\rho_i(r)=\frac{\gamma_i}{2\pi^2}\int dp\, p^2 \left(f_{i+}(\mathbf r, \mathbf p)-f_{i-}(\mathbf r, \mathbf p)\right) ~~,
\end{equation}
where the  ground-state (equilibrium) distribution functions are defined as 
\begin{eqnarray}
f_{i\pm}(\mathbf r, \mathbf p)&=&\frac{1}{1+\exp
\left[{(\epsilon_i^*(\mathbf r, \mathbf p)\mp \nu_i)/T}\right] }~
,\, i=p,n, t,h, \nonumber\\
f_{i\pm}(\mathbf r, \mathbf p)&=&\frac{1}{-1+\exp
\left[{(\epsilon_i^*(\mathbf r, \mathbf p)\mp \nu_i)/T}\right] }~
,\, i=d,\alpha, \nonumber\\
f_{e\pm}({\mathbf r},{\mathbf p})\,&=&\,\frac{1}{1+\exp[(\epsilon_e\mp
\mu_e)/T]},
\label{dist}
\end{eqnarray}
with $\epsilon_i^*(\mathbf r, \mathbf p)=\sqrt{p^2+M_i^*(\mathbf r)^2}$,
$M_i^* (\mathbf r) = M- {g_{si}}\phi_0 (\mathbf r)$, and $\epsilon_e=\sqrt{p^2+m_e^2}$.
 $\mu_e$ is the electron chemical potential, and
the nucleons and clusters effective chemical potentials, $\nu_i,\, i=p,n,d,t,h,\alpha$, are given by:
\begin{equation}
\nu_i=\mu_i - g_{vi} V_0 (\mathbf r)- {g_{\rho i}}~  t_{3 i}~ b_0 (\mathbf r)-
q_{ei} A_0(\mathbf r),
\end{equation}
where $\mu_i$ and $q_{ei}$ are, respectively, the chemical potential and electric charge of particle $i$, and $t_{3i}$ is the third component of the isospin operator.

For the entropy, we take
the one-body entropy density:
\begin{eqnarray}
 S_t&=&  -\sum_{i=n,p,t,h}\int \frac{d^3 p}{4\pi^3} ~
\left\{ f_{i+}(\mathbf r, \mathbf p) \ln f_{i+} (\mathbf r, \mathbf p) \right.
\\
&&\left. +\left[1- f_{i+}(\mathbf r, \mathbf p)\right] \ln \left[1-f_{i+} (\mathbf r, \mathbf p)\right] +
( f_{i+} \leftrightarrow f_{i-} )
\right\}.\nonumber\\
  &&  -\sum_{i=d,\alpha} \int \gamma_i\frac{d^3 p}{(2\pi)^3} ~
\left\{ f_{i+}(\mathbf r, \mathbf p) \ln f_{i+} (\mathbf r, \mathbf p)
\right. \nonumber\\
&&\left.-\left[1+ f_{i+}(\mathbf r, \mathbf p)\right] \ln
  \left[1+f_{i+} (\mathbf r, \mathbf p)\right] 
+ ( f_{i+} \leftrightarrow f_{i-} )
\right\}.\nonumber
\end{eqnarray}

 The equations of motion for the meson fields (see \cite{pasta1})  follow from the variational 
conditions:
\begin{equation}
\frac{\delta}{\delta \phi_0(\mathbf r)} \Omega =
\frac{\delta}{\delta V_0(\mathbf r)} \Omega = \frac{\delta}{\delta b_0(\mathbf r)} \Omega =   \frac{\delta}{\delta A_0(\mathbf r)} \Omega =  0 ~ ,
\label{meson}
\end{equation}
with
\begin{eqnarray}
 &&\Omega = \int_{\rm{V_{WS \, cell}}} d^3 r   \,\,\, \omega
    (\{f_{i+}\},\{f_{i-}\}, \{ F_j\}, \{\nabla F_j\}), \label{grand1}
\end{eqnarray}
where the space integral is over the volume of the Wigner Seitz cell, defined as $V_{WS}=A_i/\rho$,
 and we are using the same notion of Eq. (\ref{grand}). For
the temperatures considered in the present work, the bosonic particles,
$d$ and $\alpha$, do not condensate, so we have only considered the
thermal contributions, and did not include the condensate terms in the above expressions.

The numerical algorithm for the description of the neutral
$\{n,p,d,t,h,\alpha, e\}$ matter at finite temperature is
a generalization of the formalism presented in \cite{silvia10}.
The Poisson equation is always solved by using the appropriate Green's
function according to the  spatial dimension of interest, and the
Klein-Gordon equations are solved by expanding the meson fields in a harmonic
oscillator basis with one, two or three dimensions, based on the method
proposed in \cite{ring}. The differential equations are solved using Neumann boundary
conditions, and, when necessary, an auxiliary virtual source profile outside the cell, to
help convergence.
 One important source of numerical problems are the Fermi integrals, hence,
we have used the accurate and fast algorithm given in Ref. \cite{aparicio} for their calculations.

%%%%%%%%%%%%%%%%%%%%%%%%%%%%%%%%%%%%%%%%%%%%%%%%%%%%%%%%%%%%%%%%%%%%%%%%%
%%%%%%%%%%%%%%%%%%%%%%%%%%%%%%%%%%%%%%%%%%%%%%%%%%%%%%%%%%%%%%%%%%%%%%%
\section{Results}
\label{sec:results}

%%%%%%%%%%%%%%%%%%  
\begin{figure}[thb]
    \includegraphics[width=0.9\linewidth,angle=0]{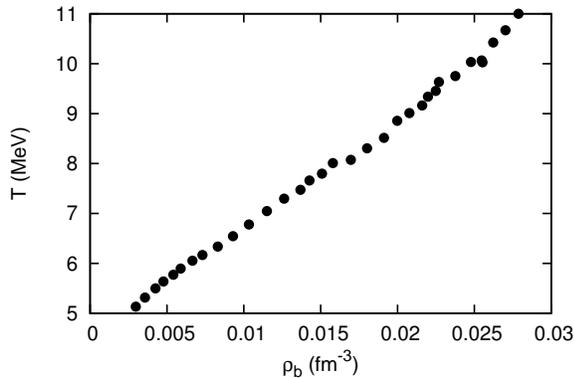}
    \caption{The range of temperatures and densities of the HIC experiment of Qin \textit{et al.} \cite{qin12}.}
\label{fig1}
\end{figure}
%%%%%%%%%%%%%%%%%%  

In the present section, we discuss how the  couplings of the light clusters to
the vector meson $\omega$ define the behavior of the equilibrium constant
$K_c$,  and the distribution of the cluster fractions. Considering the measured  equilibrium constants
(EC) \cite{qin12} as a condition for the EOS in the low-density region, 
an optimum value for the parameter $\eta$ is found. We will next calculate
the pasta phase, including light clusters, and using the same
parametrizations discussed for homogeneous matter. The effect of
including the pasta phase on the EC and the proton
and neutron chemical potential is also discussed.

\subsection{Light clusters}

%%%%%%%%%%%%%%%%%%%%
\begin{figure}[thb]
    \includegraphics[width=0.9\linewidth,angle=0]{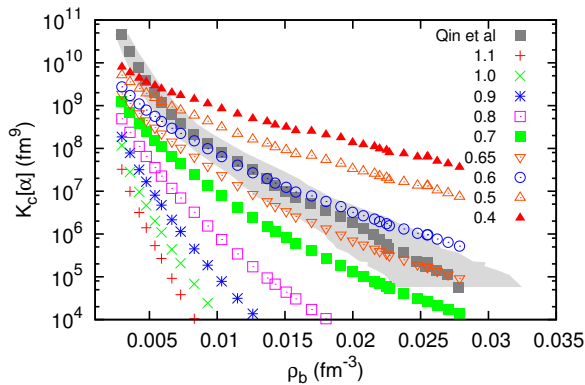}
    \caption{The chemical equilibrium constant $K_c$ for the $\alpha$-particle for different values of $\eta$.}
\label{fig2}
\end{figure}
%%%%%%%%%%%%%%%%%%%%

In Ref. \cite{qin12},  experimentally derived EC for several light clusters ($d, t, h,\alpha$) were reported. 
The range of densities and temperatures of that experiment is shown in Fig. \ref{fig1}.
In the following, we will consider these experimental observables to 
constrain the cluster coupling to the vector meson $\omega$. The chemical EC  defined in \cite{qin12} are
\begin{equation}
K_c[i]=\frac{\rho_i}{\rho_n^{N_i}\rho_p^{Z_i}},
\label{kc}
\end{equation}
where $\rho_i$ is the number density of cluster $i$ with neutron
number $N_i$,  and  proton number $Z_i$, and $\rho_p$,
$\rho_n$ are, respectively, the number densities of free protons and neutrons.
The global proton fraction, $Y_p=\sum_i Z_i \rho_i/\rho$, with $ \rho =\sum_i A_i \rho_i$,
was determined in these experiments as $Y_p=0.41$.

%%%%%%%%%%%%%%%%%%  
\begin{figure}[thb]
     \includegraphics[width=0.9\linewidth,angle=0]{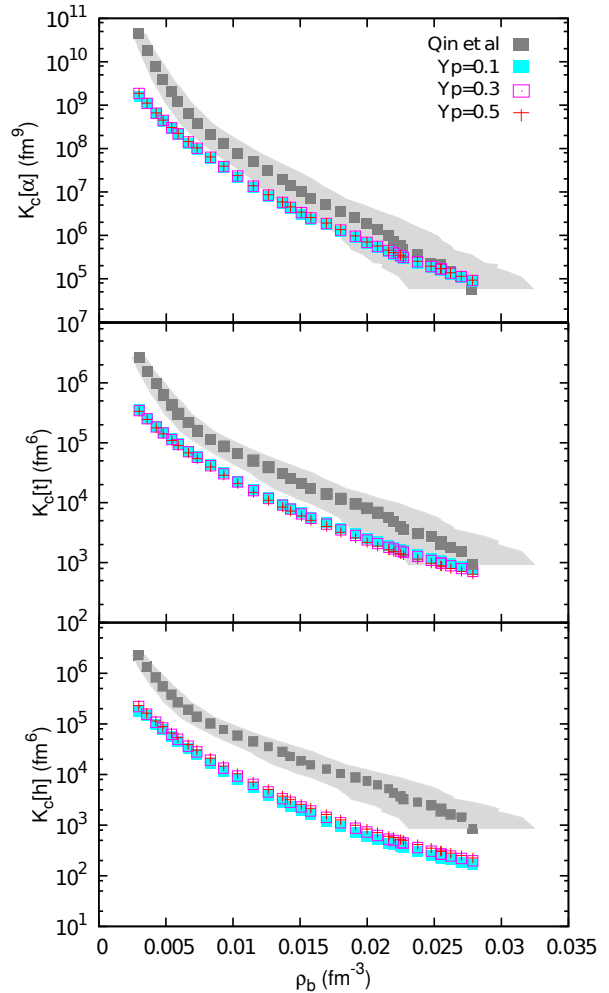}
    \caption{The chemical equilibrium constant $K_c$ for the $\alpha$-particle (top), tritium (middle), and helion (bottom) for different global proton fractions values $Y_p$, using
    $\eta=0.65$. }
\label{fig3}
\end{figure}
%%%%%%%%%%%%%%%%%%%%

We first consider the EOS for  homogeneous matter with light clusters
in chemical equilibrium, such that the chemical potential of each cluster
is given by 
$$\mu_i=N_i\mu_n+Z_i\mu_p.$$
In our calculation, a gas of protons, neutrons and light clusters is
considered in thermodynamical equilibrium.
Taking the cluster-meson parametrization proposed in the previous
section with the cluster-vector meson couplings defined in
(\ref{gveta}), we  consider $\eta$ a free parameter that will 
fix the cluster-vector meson coupling.  
We calculate the $\alpha$-equilibrium constant, $K_c[\alpha]$, for different values of
$\eta$ and plot them in Fig. \ref{fig2}, together with the
experimental results of \cite{qin12}. The calculation was performed
for $Y_p=0.41$. Taking $\eta=1$, the $\alpha$-equilibrium constant is
too small, indicating that the
parametrization is too repulsive, already for the lowest densities
considered. The experimental results seem to indicate that $\eta\sim
0.65 - 0.7$. In the following, we will consider these two values of $\eta$,
and we will discuss the cluster fraction also when heavy clusters are
included. In accordance with other models that include the interaction
between nucleons and nuclei in the low density region discussed in Refs. \cite{qin12,hempel2015},
we can reproduce the data obtained from the laboratory test of the EOS.
The deviations to the EOS at very low densities, reported also by the other approaches,
are possibly caused by the experimental difficulties to produce a state in thermodynamical
equilibrium at such densities.

%%%%%%%%%%%%%%%%%%  
\begin{figure}[!htbp]
    \includegraphics[width=0.9\linewidth,angle=0]{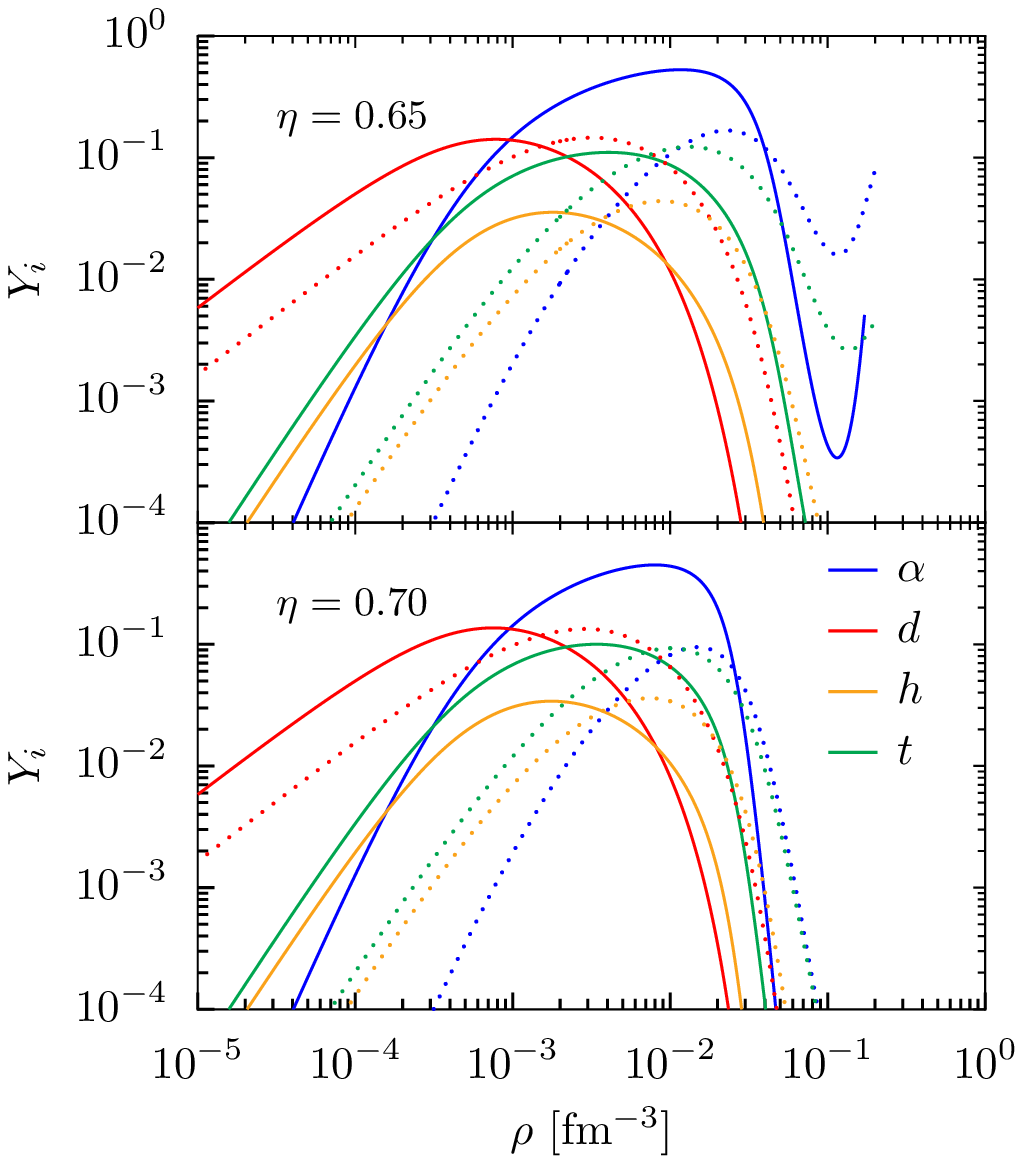}\\
    \caption{The $i$-cluster particle fraction $Y_i$ with $\eta=0.65$
      (top panel) and  $\eta=0.70$ (bottom panel)
    for different values of temperature: $5$ MeV (solid), and $10$ MeV (dotted).}
\label{fig4}
\end{figure}
%%%%%%%%%%%%%%%%%%%

%%%%%%%%%%%%%%%%%%  
\begin{figure*}[!htbp]
\begin{tabular}{c}
     \includegraphics[width=0.7\linewidth,angle=0]{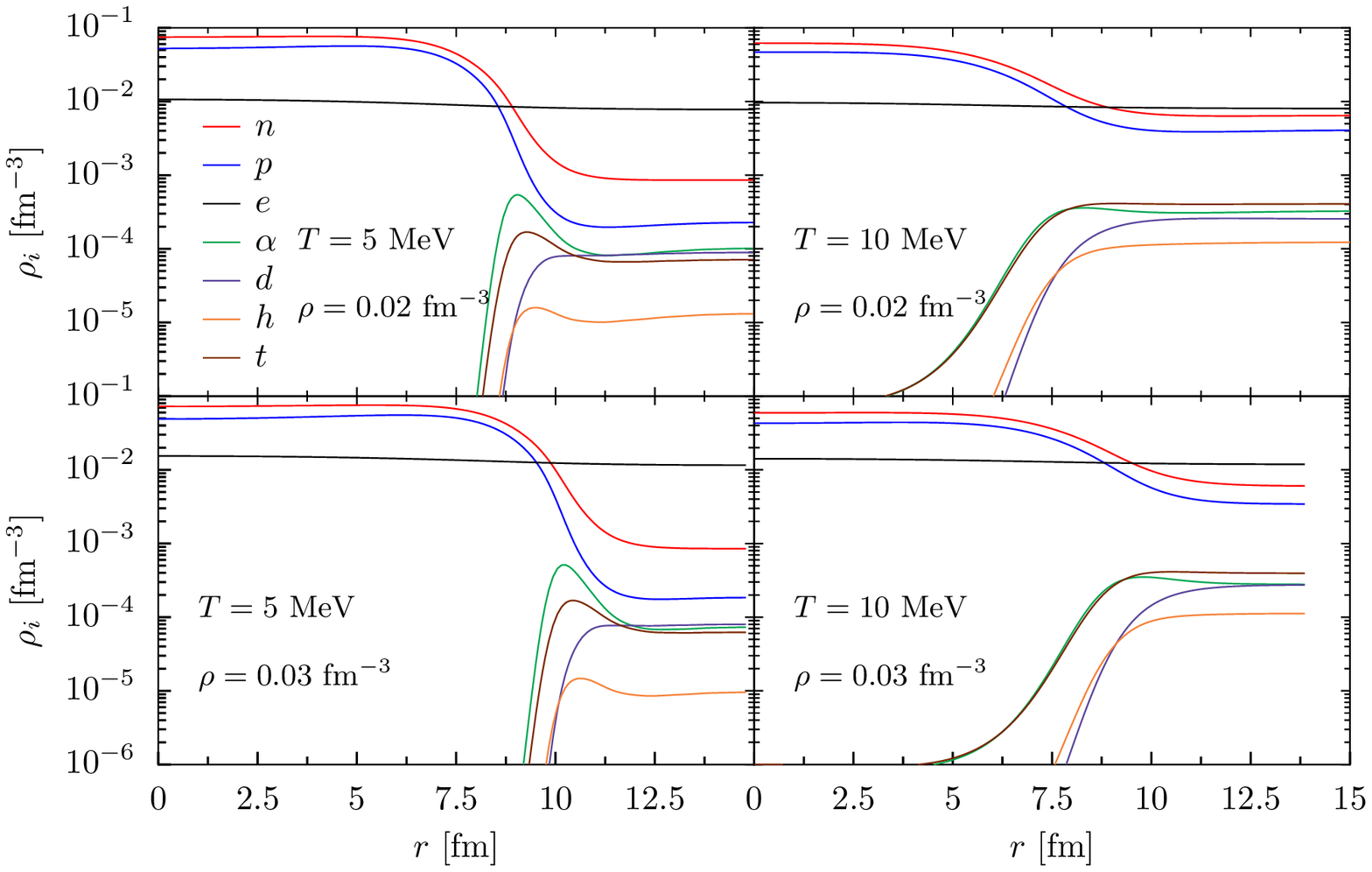}
\end{tabular}
    \caption{Density profiles for $\eta=0.7$, $Y_p=0.41$, $T=5$ MeV
    (left panels) and  $T=10$ MeV
    (right panels),  $\rho=0.02$ fm$^{-3}$ (top panels) and  $\rho=0.03$ fm$^{-3}$
      (bottom panels), obtained with the FSU parametrization.}
\label{fig5}
\end{figure*}
%%%%%%%%%%%%%%%%%%%%

%%%%%%%%%%%%%%%%%%  NEW FIG %%%%%%%%%%%%%%%%%%%%%%%%%%
\begin{figure*}[!htbp]
\begin{tabular}{c}
     \includegraphics[width=0.95\linewidth,angle=0]{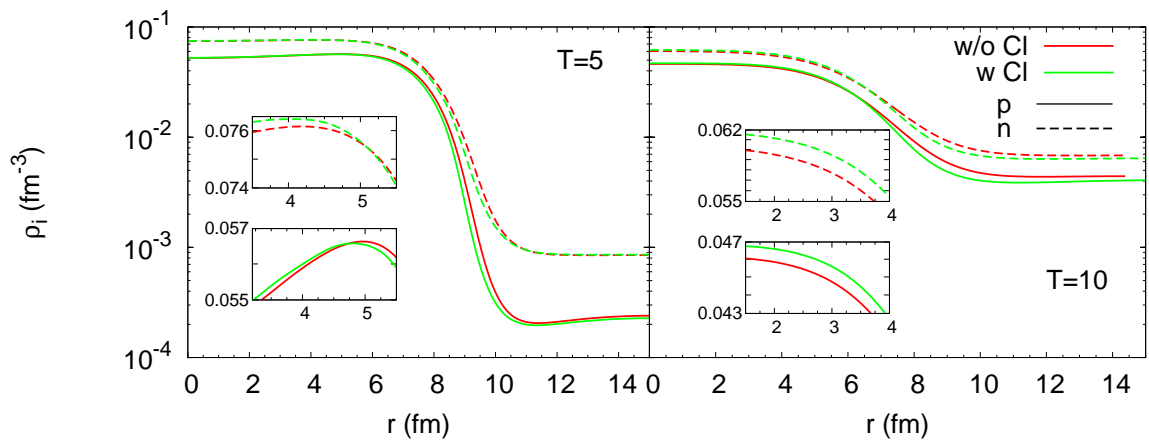}
\end{tabular}
    \caption{Neutron (dashed) and proton (solid) density profiles for $\eta=0.7$, $Y_p=0.41$, and $\rho=0.02$ fm$^{-3}$, at $T=5$ (left) and  $T=10$ MeV  (right panel), obtained for the FSU parametrization, within a TF calculation with (green) and without (red) light clusters.}
\label{fig6}
\end{figure*}
%%%%%%%%%%%%%%%%%%%%

In the following, we perform our calculations by fixing the proton
fraction to $Y_p=0.41$, which corresponds to the value extracted from the
experiment in Ref. \cite{qin12}.
However, as seen in  Fig. \ref{fig3}, the effect of
the total proton fraction on the equilibrium constant of the $\alpha$-particles is very small. It was shown in
\cite{hempel2015} that for a  non-interacting
Maxwell-Boltzmann gas  of protons, neutrons and
clusters in equilibrium, the  chemical EC do not
depend on the proton fraction. They have, however, obtained a
dependence on the proton fraction when describing matter within the 
excluded volume HS EOS \cite{HSB}, as a chemical mixture of nuclei and nucleons in
nuclear statistical equilibrium,  having the density dependent model DD2 \cite{typel10}, 
as the underlying RMF model, and accounting for the Pauli blocking 
between nucleons and nuclei in a simple approximation by using the excluded volume concept. 
 This difference was attributed to the fact
that in their calculation a gas of  interacting particles was
considered. 
The observed  relative effect in our calculation, although very small,
is, however, the same, the smaller the proton
fraction, the smaller the EC. The small dependence
on $Y_p$, close to the ideal gas result, may be
explained by the fact that the density distributions depend only on the
effective chemical potential, $\nu_i$, and the effective masses, which have
only the contribution  from the coupling to the isoscalar meson field $\sigma$.  
In the present calculation, we consider a gas of interacting
particles, as in Ref. \cite{hempel2015}, but in that work, the approach
describing the equilibrium, a nuclear statistical equilibrium formalism, is completely different, and this is
probably the reason for the different behavior.
 Contrary to the $\alpha$-clusters, the heliums and tritiums have a
non-zero isospin and, therefore, are more sensitive to the global proton
fraction of matter, as seen in Fig. \ref{fig3} in the middle and bottom
panels, although the effect of the proton fraction is still quite small. The EC changes in opposite directions since a medium with a
smaller proton fraction favors the formation of tritium and disfavors
the formation of helium, and so the  smaller $Y_P$, the larger the
tritium EC and the smaller the helium EC.

%%%%%%%%%%%%%%%%%%  
\begin{figure}[!htbp]
	\begin{tabular}{c}
     \includegraphics[width=0.9\linewidth,angle=0]{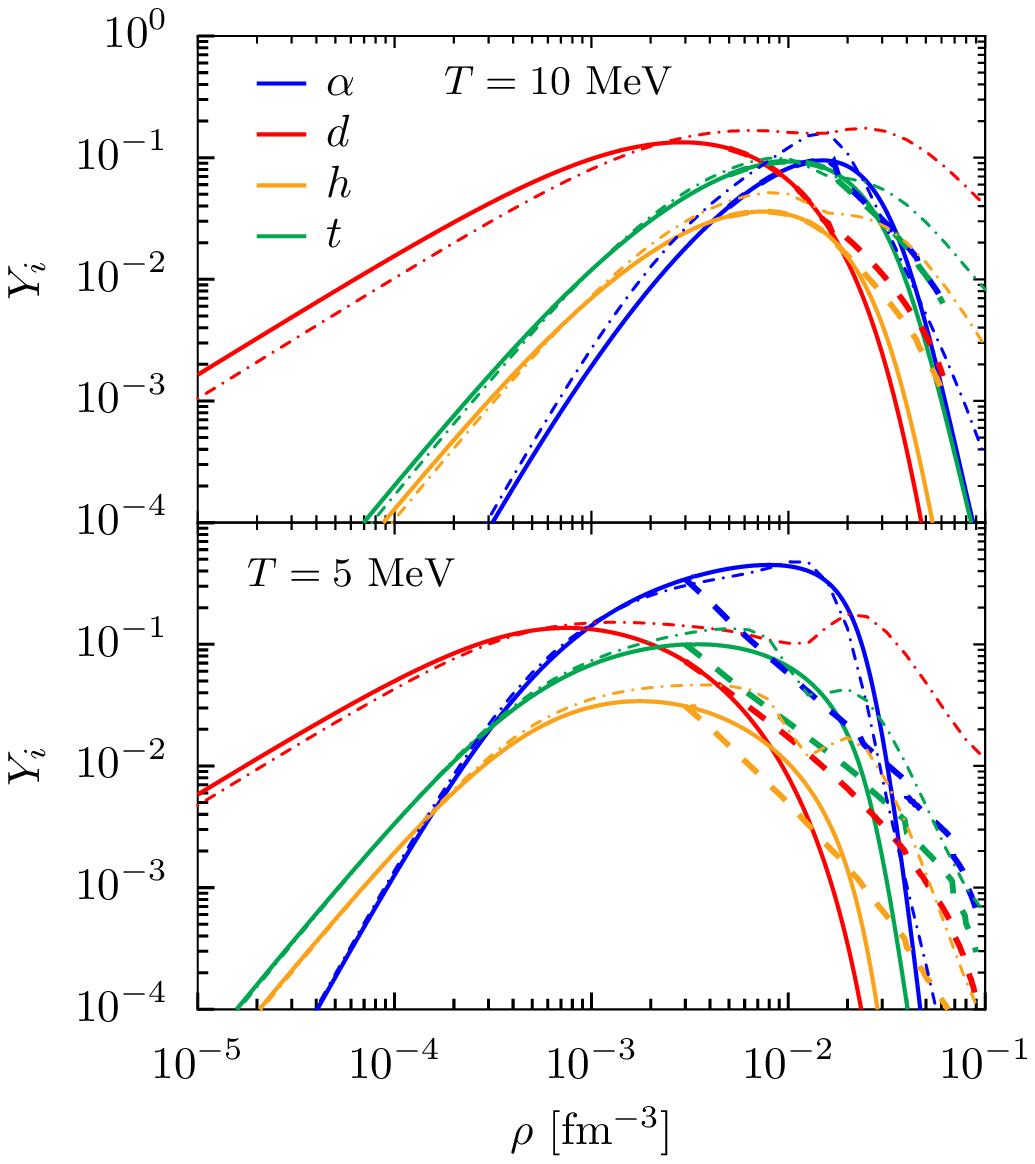}\\
    \end{tabular}
    \caption{Cluster fractions with
     $\eta=0.70$ and $Y_p=0.41$ as a function of density, 
     for $T=5$ MeV (bottom) and $T=10$ MeV (top panels). Results for a
     TF calculation (dashed), homogeneous matter with  clusters
     (solid), and the QS approach (dash-dotted lines) are shown.
       For $T=5$ MeV, the TF calculation includes the
    five geometrical configurations, droplet, rod, slab, tube and
    bubble,  for the heavy clusters. }
\label{fig7}
\end{figure}
%%%%%%%%%%%%%%%%%%%%%%%%%%%%%%%%%%%%%%%%%%%%%%%%%%%%%%%%%%%%%%%

%%%%%%%%%%%%%%%%%%  
\begin{figure*}[!htbp]
\begin{tabular}{c}
     \includegraphics[width=0.85\linewidth,angle=0]{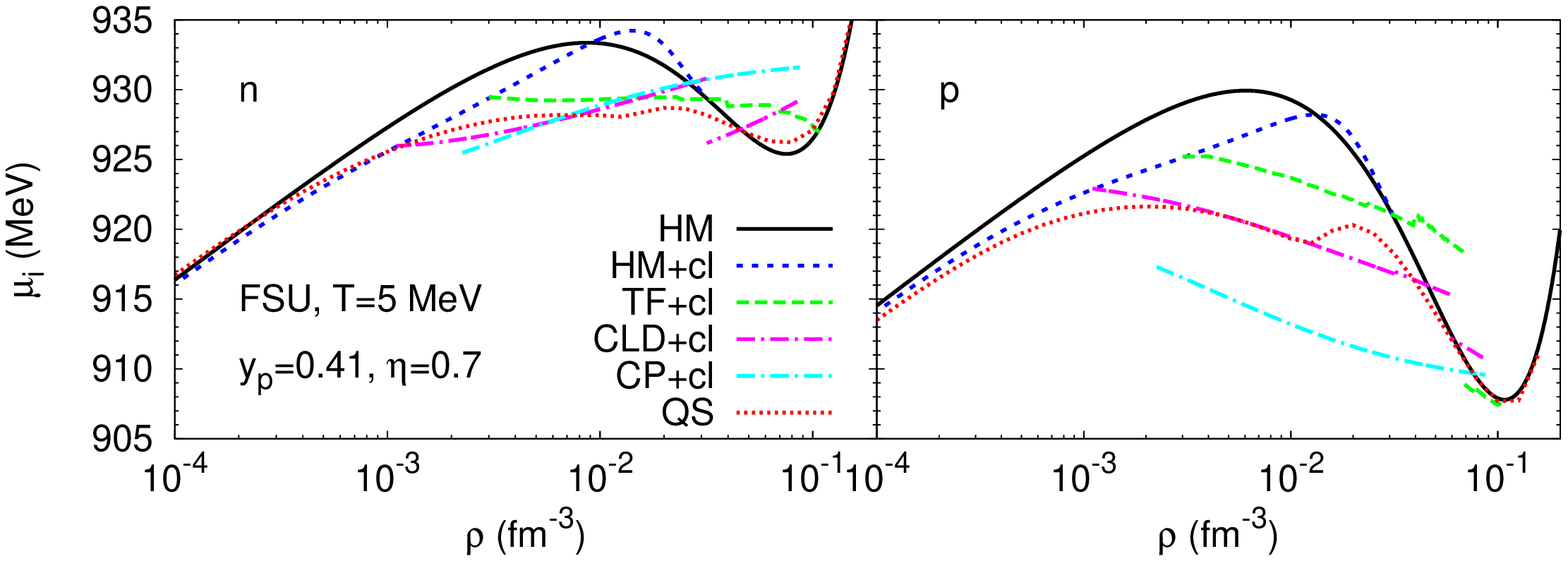}\\
      \includegraphics[width=0.85\linewidth,angle=0]{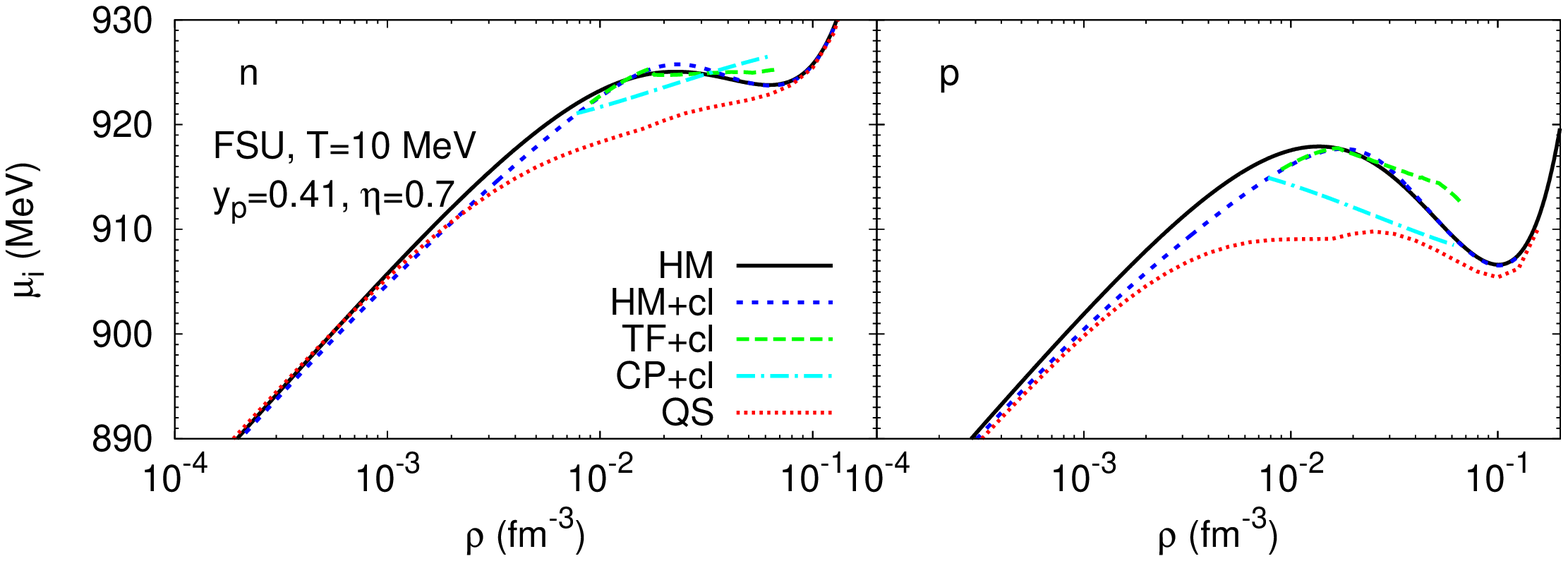}
\end{tabular}
    \caption{Neutron (left panels) and proton (right panels) chemical potentials
with $\eta=0.7$ and $Y_p=0.41$ as a function of density at $T=5$ MeV (top)
    and  $T=10$ MeV (bottom),
    for homogeneous nuclear matter (HM) (solid), nuclear matter with light
    clusters (blue short-dashed), and mean-field pasta calculations with
    clusters [TF (green, dashed), CLD (pink, dash-dotted), CP (cyan, dash-dotted)]. QS results (red, dotted) are also shown. }
\label{fig8}
\end{figure*}
%%%%%%%%%%%%%%%%%%%%

%%%%%%%%%%%%%%%%%%  NEW FIG%%%%%%%%%%%%%%%%%%%%%%%%
\begin{figure}[!htbp]
\begin{tabular}{c}
     \includegraphics[width=0.95\linewidth,angle=0]{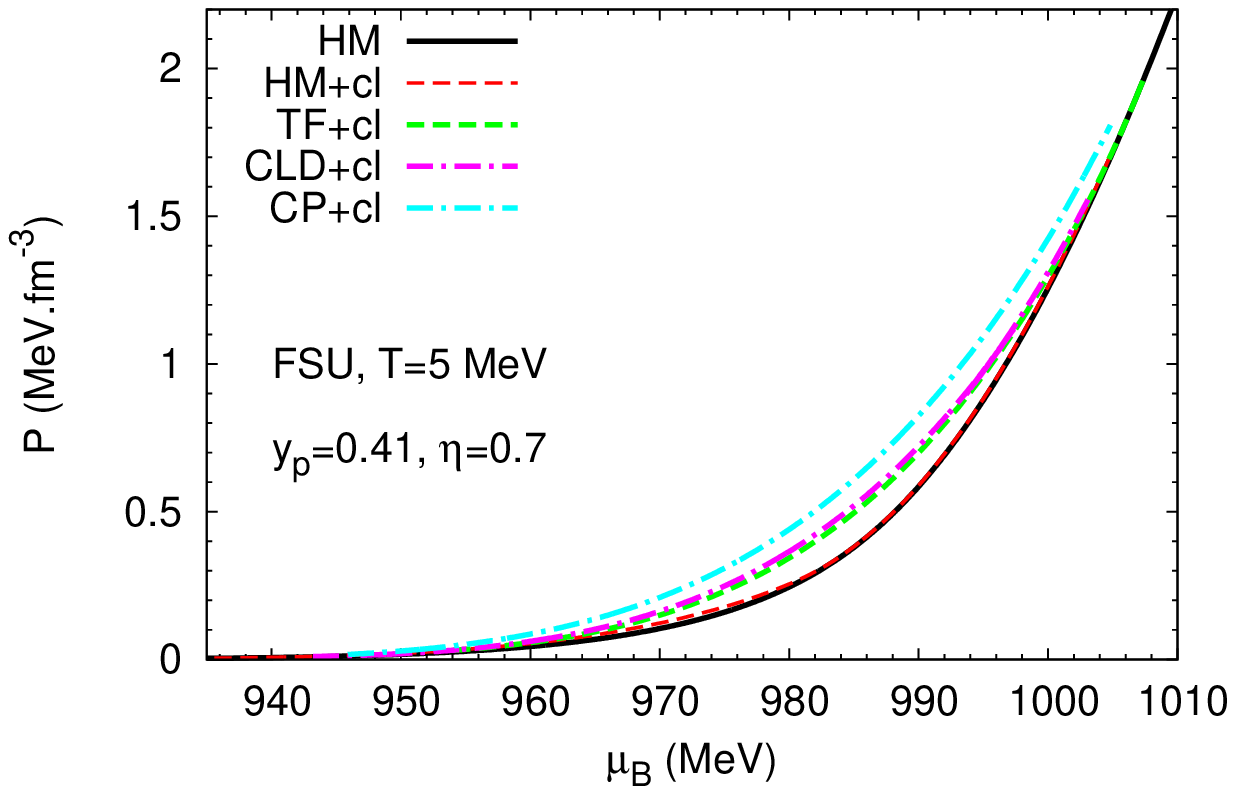}
\end{tabular}
    \caption{Pressure as a function of the baryonic chemical potential, $\mu_B$, 
  at $T=5$ MeV, for homogeneous nuclear matter (HM) (solid), homogeneous nuclear matter with light clusters (red, dashed), and mean-field pasta calculations with    clusters [TF (green, dashed), CLD (pink, dash-dotted), CP (cyan, dash-dotted)]. }
\label{fig9}
\end{figure}
%%%%%%%%%%%%%%%%%%%%
    
The fractions of the different light clusters present in homogeneous matter with
$Y_p=0.41$  are plotted in Fig. \ref{fig4},
for $T=5$ and 10 MeV, with $\eta=0.65$ (top panel)  and $\eta=0.70$
(bottom panel). Some conclusions are in order: the deuterons
are the most abundant clusters at the lowest densities due to
their smaller mass.  In fact, the relative abundance of the light
clusters at the lowest densities is mainly
driven by the fugacities $z_{i} = \exp((\mu_{i}-m_{i})/T) \approx
z_{n}^{N_{i}}z_{p}^{Z_{i}}$, and, therefore, the lightest cluster is
the most abundant at low densities; however, for $T=5$ MeV, the $\alpha$-particles become
more abundant, already below  $\rho=10^{-3}$fm$^{-3}$, due to their
large binding energy, and between $\rho=0.01-0.1$ fm$^{-3}$, they are the most abundant; the fraction of tritium is always
larger than the fraction of helion because matter with $Y_p=0.41$ is neutron rich; a larger $\eta$ reduces the
fraction of particles and moves the dissolution density to smaller
densities as expected, because a larger $\eta$ gives rise to a stronger repulsion,
induced by the vector meson; for the largest temperature, and
$\eta=0.65$, it is clearly seen that after a strong reduction of the
cluster fraction at $\rho<0.1$ fm$^{-3}$, there is a new increase  of the light
cluster fractions,
showing that the parametrization of the couplings is not repulsive
enough to dissolve the clusters at these densities. 
This is also seen for the $\alpha$-clusters at $T=5$ MeV. This problem  can be fixed by
including a mechanism that  describes excluded-volume effects, see
\cite{typel2016}, or by  introducing terms beyond a linear
dependence on the density  in the mass shifts \cite{typel10}. In the present calculation, light clusters are
considered point-like, and it is the $\omega$-meson that describes the
short distance repulsion between clusters, which, however, seems not
to be sufficient for these densities. As already mentioned above, the quasiparticle picture becomes questionable near the saturation density, and part of the correlations are already included in the mean-field approach. However, at densities of this order, other effects,
such as the formation of heavy clusters,
should be considered. This will be done in the next subsection.

\subsection{Pasta phase with light clusters}

%%%%%%%%%%%%%%%%%%%%%%%%%%%%%%%%%%%%%%%%%%%%%%%%%%%
\begin{figure*}[!htbp]
	\begin{tabular}{c}
     \includegraphics[width=0.7\linewidth,angle=0]{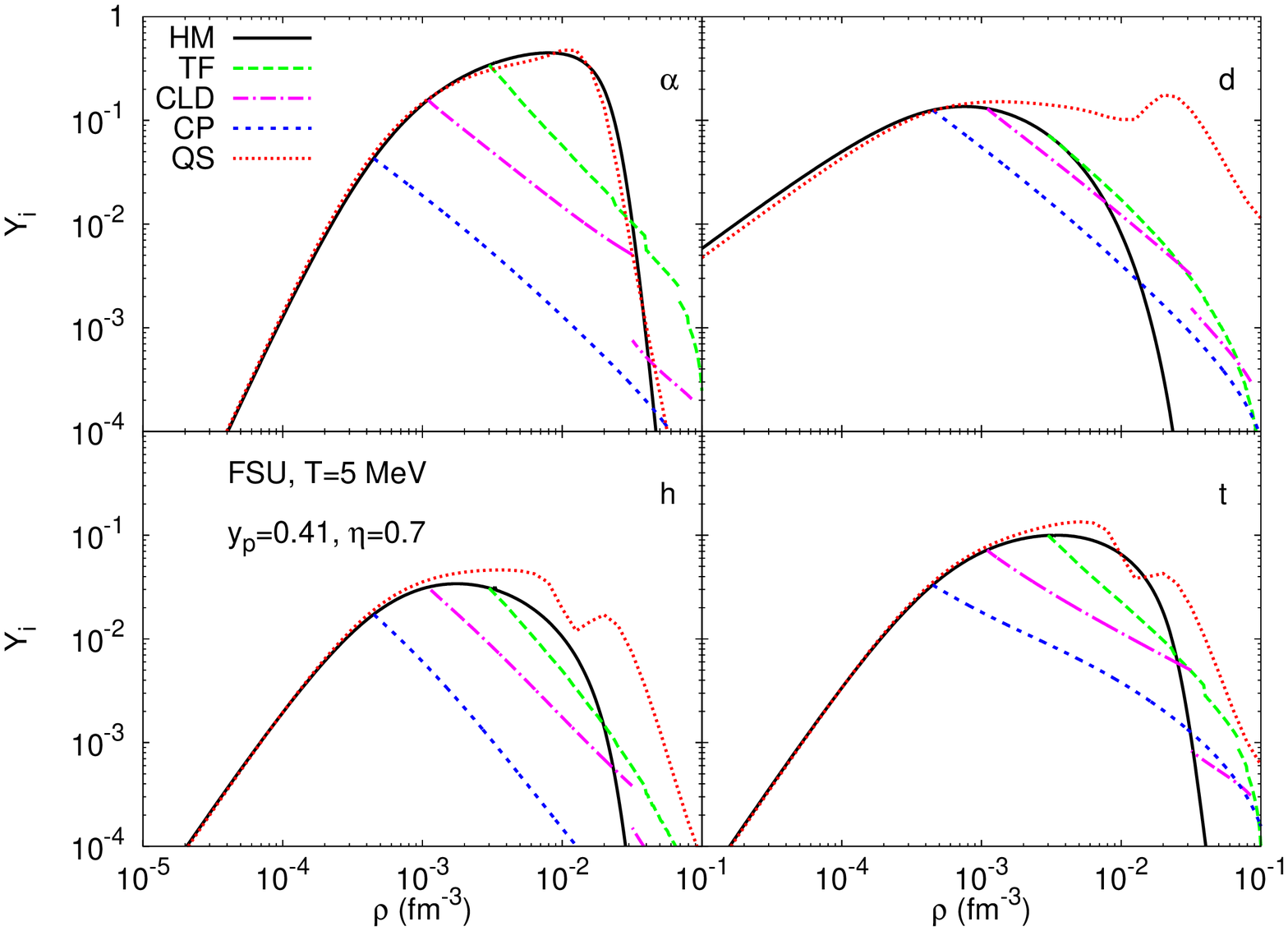}\\
    \end{tabular}
    \caption{Comparison of the cluster fractions obtained with
     $\eta=0.70$ as a function of density 
     for $T=5$ MeV:  QS calculation (red, dotted lines),
      three mean-field pasta calculations with clusters,
    TF (green, dashed), CLD (pink, dash-dotted), CP (blue,
    short-dashed), and the homogeneous matter (HM) calculation with light
    clusters (black, solid line).}
\label{fig10}
\end{figure*}
%%%%%%%%%%%%%%%%%%%%%%%%%%%%%%%%%%%%%%%%%%%%%%%%%%%%%%%%%%%%%%%%%%%%%

As seen in Fig.~\ref{fig4}, and considering $\eta=0.7$, all
  the clusters dissolve between $\rho=0.02-0.1$ fm$^{-3}$, for the two
  temperatures considered, $T=5$ and 10 MeV.  In the present subsection, we 
  choose this value of $\eta$, and we study how the appearance of
  heavy clusters is affected by the light clusters. These investigations are 
of relevance to determine the structure of the inner crust of neutron
stars, or the evolution of a core-collapse supernova matter.

We perform a Thomas-Fermi calculation,   including light clusters as degrees of
freedom, as described in the previous section. We consider the  temperatures  5 and 10 MeV, a fixed proton fraction
$Y_p=0.41$, and the cluster-meson couplings  are chosen according to Eq.
(\ref{gveta}), with
$\eta=0.7$. The calculation is performed with the FSU \cite{fsu} model
and  all geometrical configurations are considered in the calculation
with $T=5$ MeV.  For $T=10$ MeV, we only consider droplets because according
to reference  \cite{pethick1998} thermal fluctuations will
induce displacements of the rodlike and slablike clusters  which can
melt the lattice structure for temperatures  $T\gtrsim$ 7 MeV.

In Fig. \ref{fig5},  the $p, n, d, t, h$ and
$\alpha$-particle profiles for a spherical Wigner-Seitz (WS) cell at the densities
$\rho=0.02$ fm$^{-3}$ (top panels)  and $\rho=0.03$ fm$^{-3}$  (bottom
panels) are plotted.     
 For $T=5$ MeV, the light clusters present a maximum close to the
 cluster surface,  a result already
obtained in \cite{typelAIP}. Close to the surface, the largest
abundances occur  for the $\alpha$ and tritium particles, however, at the WS
cell border, the deuteron is certainly more abundant than the tritium,
and, for $\rho=0.03$ fm$^{-3}$, it even overtakes the $\alpha$-particle
density. For $T=10$ MeV, the tritium is essentially the most abundant
cluster for all densities. 
The peaked distribution at the heavy cluster surface, observed for $T=5$ MeV,  is practically washed out for all
light clusters, except for the $\alpha$-particles, when  the temperature increases.

 In order to study the effect of light clusters on the profiles
  of the heavy cluster we show, in Fig. \ref{fig6}, the $p$ and $n$
  density profiles obtained in a TF calculation with (green) and
  without (red) light clusters. In the left panel, results at $T=5$
  MeV are displayed, and in the right panel, we take $T=10$ MeV, as in
  the previous Figure. The baryonic density is set at 0.02 fm$^{-3}$
  and, at this value, the ground state heavy cluster configuration is
  the droplet, which is the geometry considered in the calculations.  
 The light clusters have a noticeable effect on the heavy cluster,
 more clearly seen at $T=10$ MeV: including clusters makes the central 
cluster  proton and neutron densities  slightly larger,  the background
gas density of both nucleons lower, the surface thickness of the
heavy cluster smaller, and the WS cell radius larger.

  \begin{table}
 \caption{
 Transition densities between the heavy clusters within a TF
 calculation with and without light clusters. The temperature is fixed
 to $T=5$ MeV and the proton fraction is set to 0.41. In the first
 column, d, r, s, t, b and HM stand for droplet, rod, slab, tube, bubble
 and homogeneous matter.} 
 \label{tab1}
   \begin{tabular}{cccc}
    \hline
    \hline
    &	no clusters & \phantom{a} & with clusters \\
     &   $\rho$ (fm$^{-3}$)& \phantom{a} & $\rho$ (fm$^{-3}$)  \\
     \hline
 d-r & 0.0230 & \phantom{a}& 0.0234 \\
 r-s & 0.0392 & \phantom{a}& 0.0396 \\
 s-t & 0.0680 & \phantom{a}& 0.0685 \\
 t-b & 0.0806 & \phantom{a}& 0.0790 \\
 b-HM & 0.101 & \phantom{a}& 0.101 \\
    \hline 
    \hline
  \end{tabular}
 \end{table} 

 In Table \ref{tab1}, we show the sequence of geometries obtained
  in a TF calculation with and without light clusters, for a
  temperature of 5 MeV and a fixed proton fraction of 0.41.  
All the five heavy cluster configurations, droplet, rod, slab, tube
and bubble,  are present in both
  calculations and the difference between the transition densities is
  small, being slightly larger when the light clusters are present,
  except for the tube-bubble transition, where it happens at a smaller
  density when no clusters are considered. 
The small effect of the light clusters on the transitions is
probably due to the fact that the largest abundances of clusters occur
for densities that favor the spherical geometry, and is below $Y_i\sim
0.005$ for all the other geometries which become stable at densities $\rho > 0.023$ fm$^{-3}$.

In Fig. \ref{fig7}, the
fraction of light clusters in homogeneous matter, solid lines, and
in the heavy-clusterized matter ({\it pasta phase}), dashed lines, are
shown. The $T=5$ MeV calculation takes into account the five clusters
geometries while for $T=10$ MeV only spherical droplets were considered. Results from a QS approach are also considered (dash-dotted lines), and they will be discussed later. At low densities, the pasta
phase calculation converges to the calculation of homogeneous matter with light clusters. This occurs below $\rho=0.001$ fm$^{-3}$, for $T=5$ MeV,
and below $\rho=0.01$ fm$^{-3}$, for the larger temperature. The presence of
the pasta clusters has two effects on the light cluster abundances:
on one hand, it reduces their abundances  for densities between
$\rho=0.001-0.01$ fm$^{-3}$, close  to the light cluster distribution maximum
in  homogeneous matter, and,
on the other hand, it extends their existence to baryonic densities well above the dissolution density of light clusters in homogeneous
matter. This behavior can be attributed  to the rather small density
of the background gas of nucleons in
a sizable fraction of the WS cell, where clusters can form abundantly.

\begin{figure*}[!htbp]
\begin{tabular}{c}
     \includegraphics[width=0.7\linewidth,angle=0]{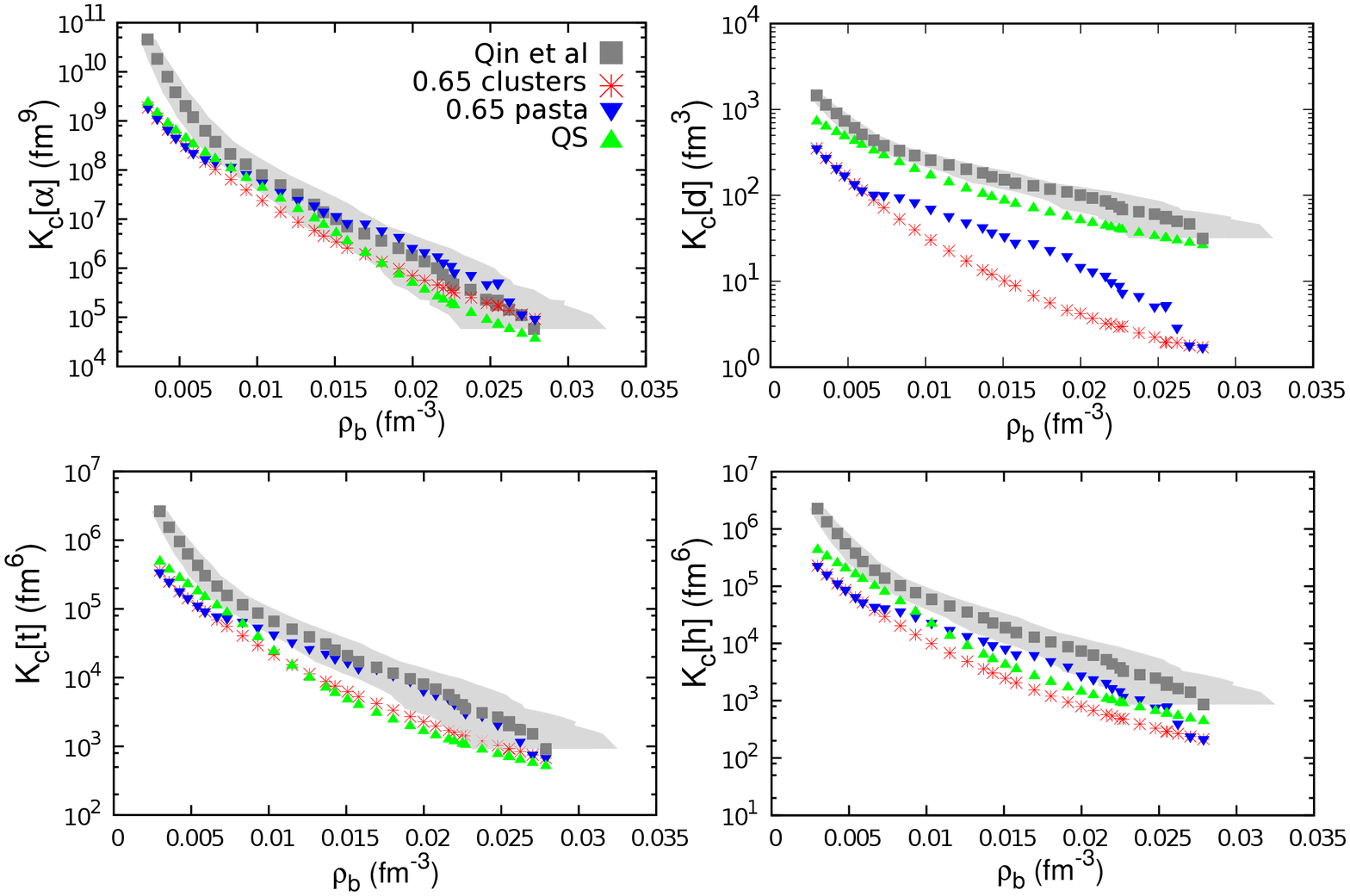}
\end{tabular}
    \caption{Chemical equilibrium constant $K_c$ for
      $\alpha$-particles (top left), deuterons (top right), tritons (bottom left), and helions (bottom right), and the parameters $\eta=0.65$, $Y_p = 0.41$, $T$, according to Fig. \ref{fig1}. }
\label{fig11}
\end{figure*}

%%%%%%%%%%%%%%%%%%%%%%%%%%%%%%%%%%%%%%%%%%%%%%%%%%%%%%%555
\begin{figure*}[!htbp]
\begin{tabular}{c}
     \includegraphics[width=0.7\linewidth,angle=0]{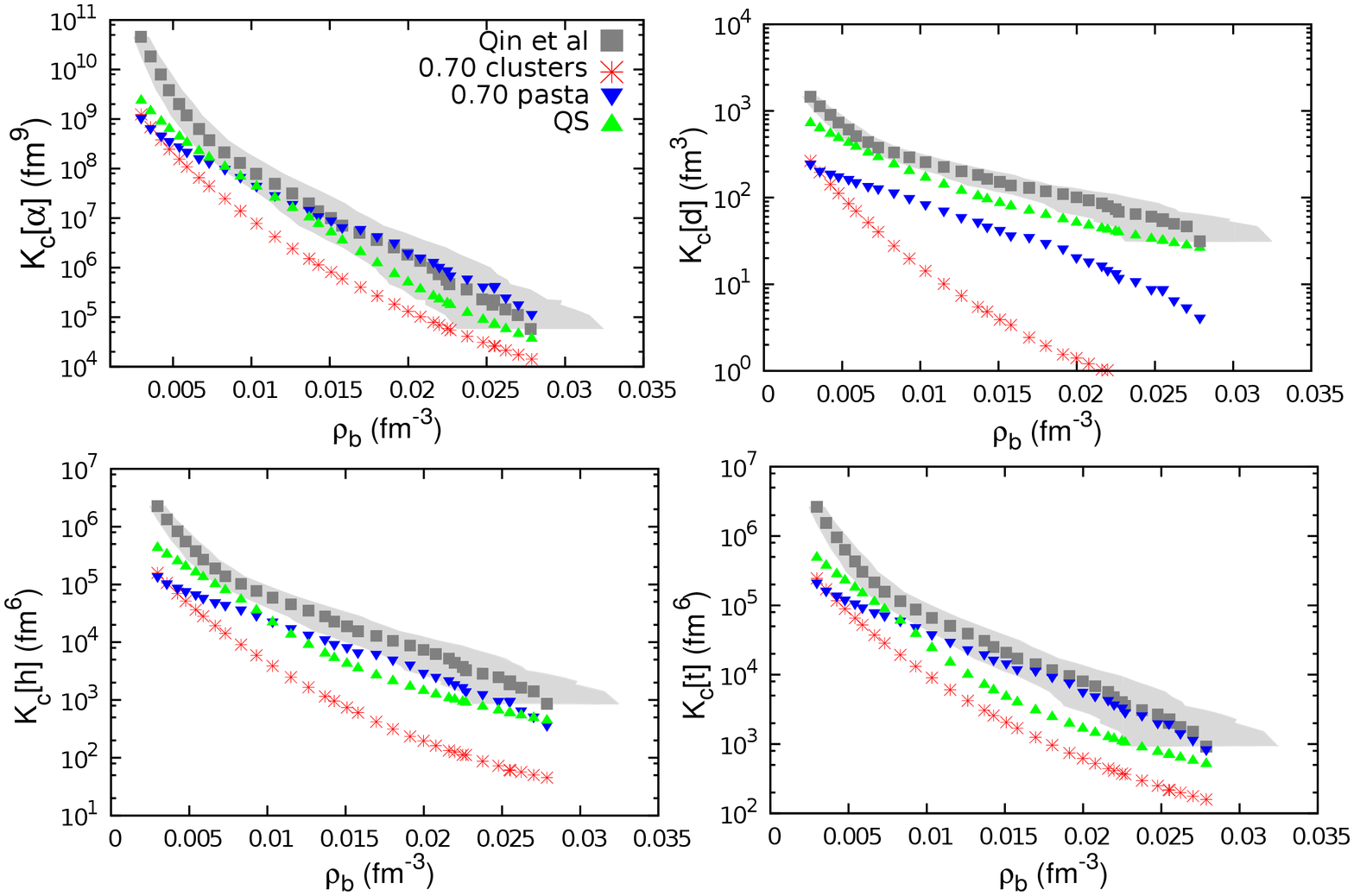}
\end{tabular}
    \caption{Chemical equilibrium constant $K_c$ for
      $\alpha$-particles (top left), deuterons (top right), tritons (bottom left), and helions (bottom right), and the parameter $\eta=0.70$, $Y_p = 0.41$, $T$, according to Fig. \ref{fig1}.}
\label{fig12}
\end{figure*}
%%%%%%%%%%%%%%%%%%%%%%%%%%%%%%%%%%%%%%%%

The effect of the presence of light clusters on the proton and neutron
chemical potentials is seen in Fig. \ref{fig8}. 
The inclusion of the pasta
contribution mainly reduces, or totally removes, the backbending that
the proton and neutron chemical potentials show for homogeneous
matter with or without light clusters, and that indicates the existence of a chemical instability. The
backbending is not totally washed out for the proton chemical
potential. However, since the calculation is done at fixed proton
fraction including electrons,  the chemical potential
that determines a possible phase transition is not $\mu_p$, but
$\mu_p+\mu_e$ \cite{hempel09}, and this one increases continuously
with density. Similar results were observed in  Ref. \cite{PCP15}, where the same FSU model was used, but a different proton fraction, $Y_p=0.3$, and temperatures, $T=4$ and 8 MeV, were
considered.
For comparison, we add to Fig. \ref {fig8} the results
  obtained within the coexistence phases (CP) approach,
  see Ref.  \cite{PCP15}, and the compressible liquid drop model (CLD)
  with clusters, see Ref. \cite{pais2017}, both calculations including light clusters. In the CP approach, the
  surface and Coulomb field contributions are added in a
  non self-consistent calculation, and, therefore, the results should
  be interpreted with caution. In particular, the approach fails
  mainly close to the transition between different phases.  This drawback
  is overtaken with the CLD model, and the transition from homogeneous
  matter with light clusters to pasta phases with light clusters is
  continuous, see the dash-dotted lines  in
  Fig. \ref{fig8}, the pink  for CLD and the cyan 
  for CP models. It is interesting that for $T=5$ MeV, CLD results are
  very similar to QS results, while TF gives larger chemical
  potentials. One of the causes of this difference is the fact that
  in the TF calculation, the electron distribution is determined
  selfconsistently, while for the CLD model it is  {\it a
    priori} taken to be constant. 

 We are also interested in investigating whether or not a
  first-order phase transition is occurring in the system. For that
  purpose,  one should look at the pressure-chemical potential
  graph. This is shown in Fig. \ref{fig9}, where we plot, for the
  different RMF approaches the pressure as a function of the baryonic
  chemical potential, $\mu_B$, which is defined as
  $\mu_B=(1-y_p)\mu_n+y_p(\mu_p+\mu_e)$, because we are considering a
  fixed proton fraction \cite{hempel09}. We include the mean-field
  pasta calculations with light clusters (TF, dashed green, CLD,
  dot-dashed pink, and CP, dot-dashed cyan) at $T=5$ MeV. The
  homogeneous matter results are given by a solid black line, and by a red dashed line, when including light clusters. We
  observe that the CP shows a jump when the transition to homogeneous
  matter occurs, which was already discussed in \cite{PCP15} and was
  attributed to the simplified treatment of the surface energy. The
  other calculations show a smooth transition. We conclude that clusterized matter has a
larger pressure at a given density, and, therefore, is more stable
than homogeneous matter, and
no first order phase
transition is expected in these range of densities.

 In order to
  understand how the mean field approach influences the light cluster fraction, we
present in Fig. \ref{fig10} a comparison of the $d,\, t,
\, h$ and $\alpha$ fractions, obtained within the five approaches, three
mean-field pasta calculation with light clusters (TF, CLD and CP) and
two calculation without the pasta structures, QS and mean-field,   for
$T=5$ MeV.  Just comparing the mean-field approaches, we conclude: TF
predicts the largest amount of light clusters,
although for the deuteron, the CLD model gives similar fractions;
 the CP model predicts fractions that are 1-2 orders of magnitude
 smaller; all pasta calculations predict the dissolution of light
 clusters at larger densities than the calculation without pasta
 structures; 
except for the $\alpha$ clusters, the QS calculations
 predict the largest amounts of light clusters at densities close to
 0.1 fm$^{-3}$ and above, however this calculation does not consider
 the possibility of heavy cluster formation. More details about these results will be
 discussed in the next section.  The CLD approach presents some
   discontinuities that are connected with the change of geometry of
   the heavy cluster,  being a limitation of considering only some
   geometries and of the single-nucleus approximation. A smoother change would be
obtained if, e.g., a full distribution of heavy clusters was
considered, and intermediate geometries are taken into account.

\subsection{Comparison with other results}

In this subsection, we continue to discuss the results of the previous subsections with respect to the
experimental data of Ref. \cite{qin12}, and compare with the many-body theoretical
calculations of Ref. \cite{roepke15}. In particular, we are interested in
understanding the effect of including the pasta phase in the
calculation of the chemical EC  
and proton and neutron chemical potentials. The experimental  chemical EC 
from \cite{qin12} have, however, to be taken with caution due to the
uncertainties on the extraction of the density and temperature from an
expanding source, since the experimental analysis is performed
considering that  thermal and chemical equilibrium was attained at the freeze-out point.

\subsubsection{Experimental equilibrium constants}

 Until now, we have focused on the chemical EC for the $\alpha$ particles. Here we discuss also the other light elements ($d$, $t$, $h$). In Figs. \ref{fig11} and  \ref{fig12}, the chemical EC  
 as defined in (\ref{kc}), calculated for
homogeneous matter with light clusters (red crosses), and pasta with light clusters (blue triangles) are plotted
together with the experimental results of \cite{qin12}  and the
results obtained within a many-body quantum statistical approach
\cite{roepke15}. It has been discussed in  \cite{hempel2015}  that the
comparison with experimental data should be performed only
considering light clusters, with $Z\le 2$, since these particles
evaporate from a relatively small source and very small quantities of
$^6$Li and $^7$Li are detected. We will consider both the calculation
including light clusters in a gas of a homogeneous distribution of
protons and neutrons, and in a pasta phase calculation. In this case,
the heavy clusters are represented by a single heavy cluster,
generally known as single nucleus approximation (SNA).

The curves obtained for the $\alpha$-particle chemical EC 
 for homogeneous matter with light
clusters are always below the experimental data, within the
uncertainty of the experimental analysis for $\eta=0.65$ or a bit
below for $\eta=0.7$, in accordance with 
Fig. \ref{fig2}. The same trend is obtained for the other three
light clusters, $t,\, h,\,$ and  $d$. The inclusion of the heavy
clusters in the calculation brings the chemical EC  
closer to the experimental results. A similar effect was shown in
\cite{hempel2015} for the STOS EOS of Shen {\it et al.} \cite{stos} and
the LS EOS of Lattimer and Swesty  \cite{LS}: the EC determined including 
heavy clusters are larger.  As in LS and differing from STOS, our
results with 
the heavy clusters agree with the experimental EC,
while the calculation obtained considering only light clusters
originates too small EC. We consider that this may be due to the fact
that 
the number of light clusters with respect to the free nucleons for a
given density is larger, the larger the different number of cluster species are
taken into account. This behavior has been presented in
\cite{hempel2015}, where, using the  EOS of Hempel and Schaffner-Bielich (HS) \cite{HSB}, the calculation of the EC was
carried out considering $np\alpha$ matter, as well as  matter with $A\le 4$,
$A\le10$ and no restriction on $A$. The larger the number of particles
included, the larger the EC obtained.

 While the
results for the $\alpha$,  $t$ and even the $h$ particles are consistent with the
experimental results, the deuteron EC 
are too low, and not even the inclusion of the heavy clusters is enough to reproduce
the experimental results.  In the present approach, the
coupling of mesons to the light clusters mimic the many-body effects
that give rise to the formation of clusters. In fact, as discussed in
\cite{roepke15}, medium modifications due to self-energies and Pauli
blocking effects prevent the use of a simple picture that considers  the chemical
equilibrium of free nuclei.  The in-medium effects are  included in our mean-field
description through  an appropriate choice of the coupling
constants of the mesons to the light clusters. It is expected that the
heavier clusters may be reasonably described, but the smaller the
cluster the more important the quantum statistical effects are. Deuterons,
being the lightest clusters will, therefore, be more sensitive to the
approach and, in order to be realistically described, a more
fundamental formalism is required \cite{roepke09,typel10,roepke11,roepke15}.

%%%%%%%%%%%%%%%%%%%%%%%%%%%%%%%%%%%%
\begin{figure}[thb]
\begin{tabular}{c}
     \includegraphics[width=0.75\linewidth,angle=0]{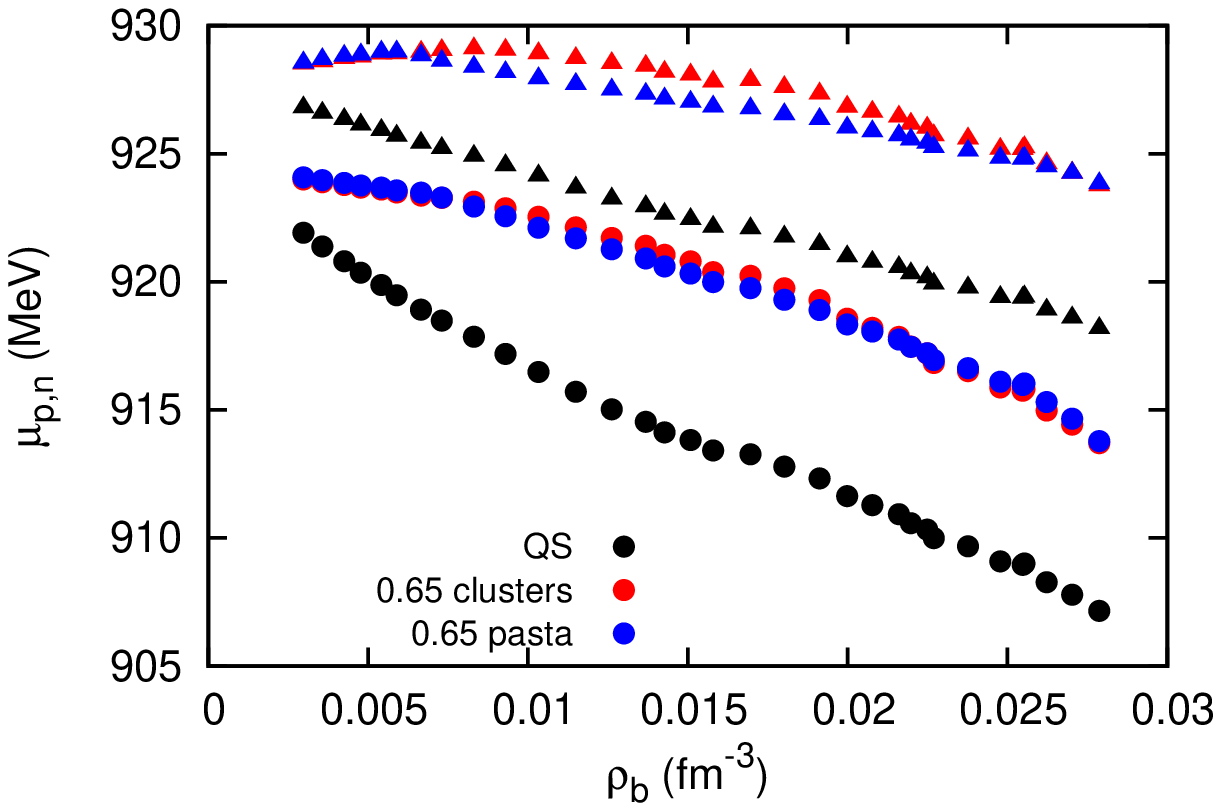} \\ 
     \includegraphics[width=0.75\linewidth,angle=0]{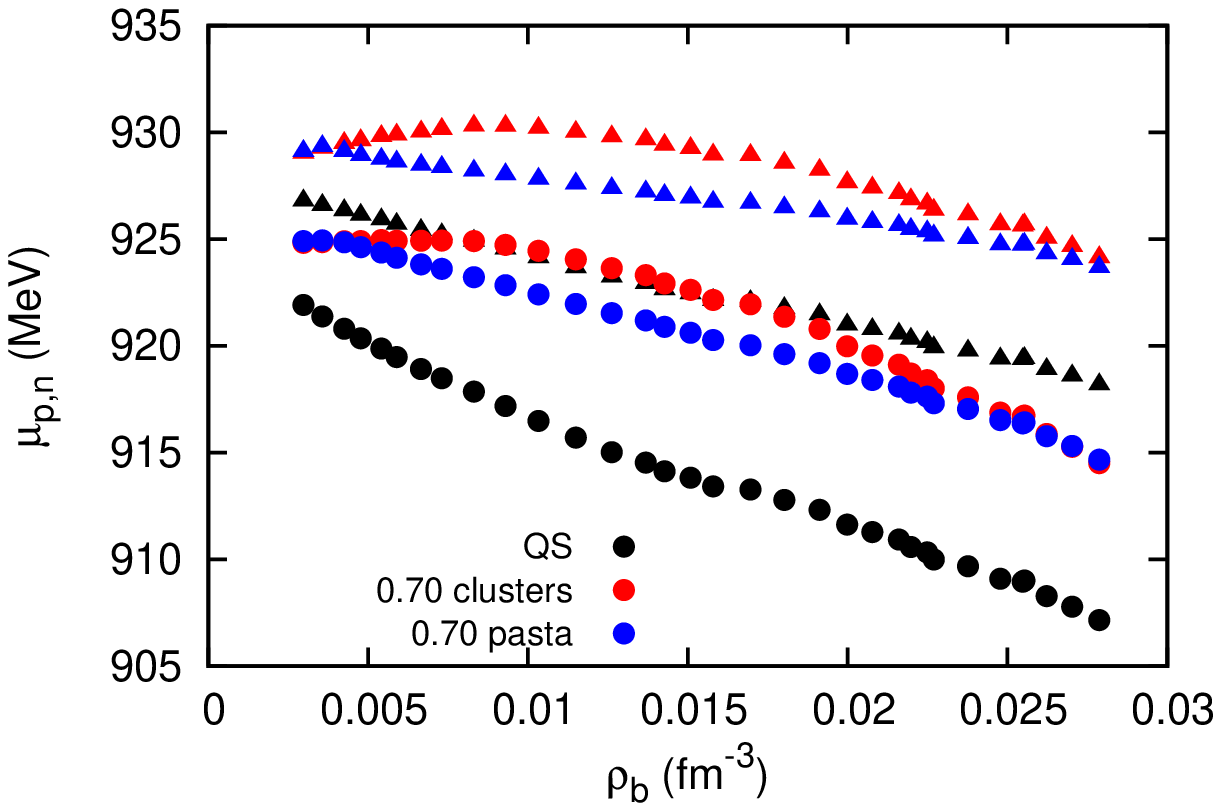}\\
\end{tabular}
    \caption{Proton chemical potential, $\mu_p$, (circles), and neutron chemical potential, $\mu_n$, (triangles), for the parameter $\eta=0.65$ (top), 0.70 (bottom panel), and $Y_p=0.41$, for homogeneous matter including light clusters (red), pasta calculation (blue), and the QS approach (black). }
\label{fig13}
\end{figure}
%%%%%%%%%%%%%%%%%%%%

\subsubsection{Quantum statistical results}

In order to compare the predictions of our mean-field approach with
the more fundamental many-body quantum statistical (QS) description
\cite{roepke15}, we have also included in Figs.  \ref{fig7} -
\ref{fig13} the corresponding QS results.  It should, however, be
  stressed that contrary to the RMF approach just discussed, the
  present QS results do not have the contribution of heavy clusters.
Starting with Figs. \ref{fig11} and \ref{fig12}, we see that the experimental EC data are well described by the QS calculations. Note that small deviations from the results given in \cite{hempel2015} are caused by the use of the more recent  expressions for the momentum dependent shifts, given in \cite{roepke15}. A reasonable agreement with the RMF model, including light cluster formation, is also obtained for the parameter $\eta = 0.7$ and the account of pasta  formation. A comparison of the cluster fractions, calculated from the different models in a wide density region, is shown in Figs. \ref{fig7} and \ref{fig8},  where we have plotted, for $\eta=0.7$, the particle
fractions at $T=5$ and $10$ MeV, with and without pasta
clusters, including also the corresponding results obtained within a QS calculation.

At low densities, in Fig. \ref{fig7}, the particle fractions for the QS and RMF results agree
quite well, except for a over production of deuterons in RMF
with respect to QS. This difference increases with temperature. The reason is clear:
we have to take into account also the continuum contribution \cite{SRS,HS}
to obtain the correct virial expansion. The contribution of continuum correlations 
to the EOS \cite{roepke15} is increasing with increasing temperature.
An approach to combine both, the virial expansion and the RMF theory, was given
in \cite{VT}.

The agreement of both calculations of cluster fractions stops at the maximum of the
particle distributions. At larger densities, particle fractions are
generally larger within the QS description. In this QS calculation, a
momentum dependent Pauli blocking was implemented and the larger the
momentum the  weaker the Pauli blocking,  originating larger mass fractions of light clusters \cite{roepke15}.  The relatively large contribution of cluster fractions from the QS calculation, especially the two-nucleon correlations ($d$) near the saturation density, is also seen in Fig. \ref{fig10}.

At larger densities, heavy clusters will  form and, contrary to the RMF
calculation, these have not been
considered in the QS calculation. It is precisely at the densities
where the momentum dependence of QS  Pauli blocking effect is more
strongly felt that the heavy clusters appear. The inclusion of the pasta phases postpones the dissolution of light clusters to larger densities but also reduces their relative mass fractions. The RMF
calculation includes the backreaction of the light clusters on the
mean-field, an effect that is not taken into account in the QS
calculation which,  therefore, predicts too strong correlations at
larger densities.  Correlations, in particular two-nucleon correlations, are present in nuclear matter also near the saturation density, but are included in the effective mean field which is fitted to the properties of dense nuclear matter.

In Fig. \ref{fig8}, we include the QS proton and neutron
chemical potentials together with the RMF ones.  These quantities
indicate that the correlations included within each approach are
different and stronger in the QS calculation. At low densities, both approaches agree reasonably well, in
particular for neutrons. The inclusion of the heavy clusters lowers the chemical
potential, becoming closer to the QS values.  The backbending effect on
the neutron and proton chemical potentials in homogeneous matter with
no clusters is the signature of an instability that originates a
liquid-gas like phase transition. In the QS approach, this backbending is
reduced but not totally removed, and, therefore, the liquid-gas
instability is still present,  indicating that heavier clusters must be
considered, see \cite{raduta10}. The RMF pasta phase calculation removes
the backbending of the neutron chemical potential, and reduces a lot
the backbending effect of the proton chemical potential. The remaining
effect is removed by the electron contribution that has also been
included in the calculation to neutralize matter.

 In Fig. \ref{fig13}, we have plotted the proton and neutron
 chemical potentials for the densities and temperatures at which the
 EC are measured. We consider both the RMF results with and without
 pasta for $\eta=0.65$, $\eta=0.7$, and the  QS results. The densities
 and temperatures tested correspond precisely to the range of
 densities where the larger discrepancies between the chemical
 potentials calculated within each framework differ the most. 
 The inclusion  of the pasta lowers the chemical potential, as shown before, but the
 chemical potential still remains essentially 5 MeV larger than the
 QS results. In contrast to the EC results where both approaches, 
 the RMF as well as the QS approach, reproduce reasonably well the 
 measured data, the results for the chemical potentials $\mu_n,\mu_p$
 are quite different. The chemical potentials contain the single-nucleon mean-field shifts which are depending on the RMF parametrization, in our case the FSU model \cite{fsu} for the pasta calculation including light clusters and the DD2-RMF \cite{typel10} used in the QS calculations. Calculating the cluster fractions, these mean-field shifts compensate nearly. Consequently, the EC, see  Figs. \ref{fig11} and \ref{fig12}, show a good agreement between both approaches.

\section{Conclusions} \label{sec:conclusions}

In the present study, we have calculated the equation of state at  low
density, including  light clusters with $A\le4$ as new degrees of freedom, besides
protons and neutrons, within three different RMF calculations:  
the Thomas-Fermi, the coexistence phase, and the compressible liquid drop models. Results from a quantum statistical calculation were also discussed, for comparison.  We have considered two different scenarios: a) the
light clusters are in equilibrium with an homogeneous distribution of
protons and neutrons; b) the nucleons clusterize and the light
clusters coexist with a heavy cluster and a proton-neutron background
gas. It has been shown that including heavy clusters shifts the light
cluster Mott densities to larger values, although reducing the mass
fraction of each type.

The RMF description of light clusters requires a reasonable choice
of the cluster-meson couplings. This has been implemented considering
both many-body quantum statistical calculations and experimental
results from HIC. Compared with the results shown in Ref. \cite{PCP15}, 
the introduction of the parameter $\eta$ which parametrizes the interaction of the meson fields
with the light clusters, allows  a reasonable description of the measured EC data.
With respect to the chemical potentials, $\mu_n$ and $\mu_p$, larger deviations are obtained, 
if comparing with QS calculations. These QS calculations will be modified
taking the formation of pasta structures into account.

 Through the coupling of the light clusters to the mesons,
we expect to take into account the  backreaction effect of the clusters in the
medium, together with an effective description of the in-medium
particle self-energies and Pauli blocking effects. For densities below $0.1 n_{sat}$, good agreement between the different approaches is obtained for the cluster fractions.  However, further
studies should be carried out to understand how the many-body effects
can be effectively taken into account if the region of higher densities, near the saturation density, is investigated. If attractive correlations are included through too strong couplings, 
it  may occur that light clusters will not dissolve at large densities, and the appropriate treatment of correlations, for instance within a density functional formalism, has to be worked out.

The simultaneous treatment of light clusters and pasta phases in warm and dense nuclear matter is of substantial relevance for various applications in HIC and astrophysics. As an example, we refer to the structure of neutron stars. The clusterization of the background gas in the inner crust
has certainly important effects on the transport properties. 
The fast decrease of the particle fractions just below $0.1$ fm$^{-3}$
coincides with the crust-core transition density. The presence of light clusters will affect 
the neutrino reaction and diffusion processes as well as
transport properties, such as electrical conductivity and specific heat. The present work contributes to the investigation of the state of warm and dense matter  when in addition to the formation of pasta phases, also light clusters have to be taken into account which require a quantum statistical approach.

%%%%%%%%%%%%%%%%%%%%%%%%%%%%%%%%%%%%%%%%%%%%%%%%%%%%%%%%%%%%%%%%%%%%%%%%%%%%%%%% 
%%%%%%%%%%%%%%%%%%%%%%%%%%%%%%%%%%%%%%%%%%%%%%%%%%%%%%%%%%%%%%%%%%%%%%%%%%%%%%%%
\section*{ACKNOWLEDGMENTS}

This work  is  partly  supported  by FCT (Portugal) under projects
UID/FIS/04564/2016 and  SFRH/BPD/95566/2013 (H.P.), by
``NewCompStar'', COST Action MP1304, and by CNPq. 

%%%%%%%%%%%%%%%%%%%%%%%%%%%%%%%%%%%%%%%%%%%%%%%%%%%%%%%%%%%%%%%%%%%%%%%%%%%%%%%
%%%%%%%%%%%%%%%%%%%%%%%%%%%%%%%%%%%%%%%%%%%%%%%%%%%%%%%%%%%%%%%%%%%%%%%%%%%%%%%%

\end{document}